# The Vavilov-Cherenkov radiation in Abraham's theory [1)]

Sergey G. Chefranov

Obukhov Institute of Atmospheric Physics RAS Moscow
schefranov@mail.ru

## Abstract

It is shown that the threshold of the observed Vavilov-Cherenkov radiation (VCR) is more accurately described by the new quantum theory of VCR (PRL 2004), based on the theory of Abraham (1909), compared with the quantum theory of VCR Ginzburg (1940), based on the theory of Minkowski (1908). Only due to the use of Abraham's theory, the characteristic microscopic mechanism of the emission of a quantum of electromagnetic radiation by a shock-polarized medium, rather than by a charged particle directly, is taken into account. The new theory of the VCR provides a basis for revising the well-known argument about the "usefulness" of applying only Minkowski's theory in the quantum theory of the VCR, used in discussing the age-old dilemma of the theories of Minkowski and Abraham, which alternatively describing the electromagnetic field (EMF) in the medium. This discussion continues to this day, despite the fact that Minkowski's theory has been rejected by experiments since 1975 to directly measure Abraham's force, which complements the Lorentz force and is absent from Minkowski's theory at any EMF frequencies. The application of the new theory of VCR to describe the VCR effect in systems for which the well-known theory of VCR is inapplicable (Bessel rays, isotropic plasma) is considered.

## 1. Introduction

The fundamental effect of Vavilov-Cherenkov radiation (VCR) is used, for example, to identify particles according to Cherenkov detectors based on the well-known theory of VCR [1]-[4]. At the same time, the problem of the accuracy of identification [2] of relativistic particles with high energy remains unresolved when the classical macroscopic VCR theory of the Tamm-Frank [1] and the quantum theory of Ginzburg (1940) [4], based on the theory of Minkowski (1908) [5], are used to interpret the observational data.

It is known that both of these theories of VCR do not take into account the microscopic mechanism of emission of VCR by a polarized medium [1]-[4], rather than by a fast charged particle directly, which was first noted by Tamm himself [1] (I. E. Tamm, 1939). As a result, this theory of the VCR gives the correct angular position only for the interference maximum in the observed angular distribution of the intensity of the VCR field already existing in the medium, but does not correspond to the maximum observed angle of the VCR cone, which determines the threshold for the realization of the VCR effect (see Fig. 1 and Table 1 in the next section).

Moreover, the Minkowski theory used in the VCR Ginzburg theory (1940)[4] was rejected in 1975 by experiments [6]-[8], which confirmed the existence of the known Abraham force [3], which is absent in Minkowski's theory and introduced in Abraham's theory (1909) [9], [10].

The new quantum theory of VCR, precisely due to the use of Abraham's theory, already takes into account the microscopic mechanism of threshold emission of VCR by the medium [11]-[13]. In [11], a relativistic generalization of the Landau superfluidity criterion was obtained, which corresponds to the mechanism of energetically favorable generation of a quantum by the medium in the entire observed range of angles of the cone by analogy with the anomalous Doppler effect and the emission of negative energy waves in flow systems. There is a better agreement with the observational data of the VCR threshold than for the theories of Ginzburg and Tamm-Frank (see See table 1 in the next section).

The description of the threshold emission of radiation by Bessel rays is considered and in other cases inaccessible to the VCR theories by Tamm-Frank and Ginzburg, such as the emission of transverse microwave waves in an isotropic plasma is provided in [11].

The example obtained in [11]-[13] of the application of Abraham's theory in the theory of quantum mechanics indicates the unjustification of the popular idea of the "usefulness" of using Minkowski theory in the quantum VCR theory. This idea still exists, despite the fact that Minkowski's theory was rejected by experiments to measure the Abraham force. All these issues are discussed in more detail in the following sections. The following section 2 supplements this Introduction and is also introductory in nature.

## 2. Comparison of the VCR theories based on the Minkowski and Abraham's theories

When a charged particle moves fast enough relative to the medium, it can cause the medium to coherently emit electromagnetic energy, observed in the form of Cherenkov radiation or Vavilov-Cherenkov radiation (VCR) [1]-[4].

Indeed, as noted in [1]-[4], the fact that the emission of the VCR is carried out precisely by the medium, and not by a fast charged particle directly, is a characteristic feature that distinguishes VCR from other types of radiation emitted by the moving charged particle. Therefore, for the emission of VCR radiation by a medium, a specific shock polarization of the medium is necessary, which can be carried out even by a charged particle moving uniformly at a sufficiently high speed, by analogy with a shock wave arising in the air when an aircraft is moving in it at a speed exceeding the speed of sound [1]. Such a mechanism of the emission of VCR by the medium is not related to the need to accelerate or slow down the motion of a charged particle in or near the medium.

In this regard, Tamm [1] noted (see page 79 in [1]): "From the point of view of microscopic theory, the radiation in question is not emitted directly by an electron, but is caused by coherent vibrations of the medium molecules excited by an electron."

Similarly, in the course of theoretical physics by L. D. Landau and E. M. Lifshits [3] (see paragraph 86 on page 448 in [3]), as an opposit to the bremsstrahlung emission of electron, the following statement is made: "In the Cherenkov phenomenon, we are dealing with radiation emitted by a medium under the influence of a field moving in it particles".

The VCR effect is widely used in detectors and particle counters, as well as for the generation of laser radiation, for example, in photonic crystals [14]-[20].

In all studies based on the VCR observational data, to interpret these data until now, only the well-known VCR theory [21]-[25] is used, which, however, does not explicitly take into account the specified mechanism of the VCR emission by a polarized medium.

Indeed, in both the macroscopic Tamm-Frank theory (1937)[22] and the quantum VCR theory of the Ginzburg (1940) [23] (see also [24],[25]), on the contrary, the VCR emitter is precisely the fast charged particle itself. Accordingly, in the theory of the VCR [22]-[25], the medium is not considered as a direct radiator of the VCR, and therefore the change in the energy of the medium associated with the emission of the VCR quantum is not explicitly taken into account at all.

However, in this regard, it is nevertheless stated in [4] that "over the many years of the existence of the VCR theory, there has been no doubt about its relevance in describing the observed VCR effect and in its practical application."

Indeed, in the works noted in [4] (see also the critical discussion of the known VCR theory in [26]-[30]), no alternative theory of the VCR was proposed, which would explicitly take into account the change in the energy of the medium when it emits the VCR quantum.

At the same time, in [4] it is meant that the known VCR theory [22]-[25] describes quite well the coherent field of the VCR already established in the medium and the angular position of the interference maximum corresponding to the observations at the value $\theta = \theta_C$ (marked in Fig.1 with a dotted line) in the angular distribution of the VCR intensity $I(\theta)$, which has the form [31] (see Fig.4 in [31]):

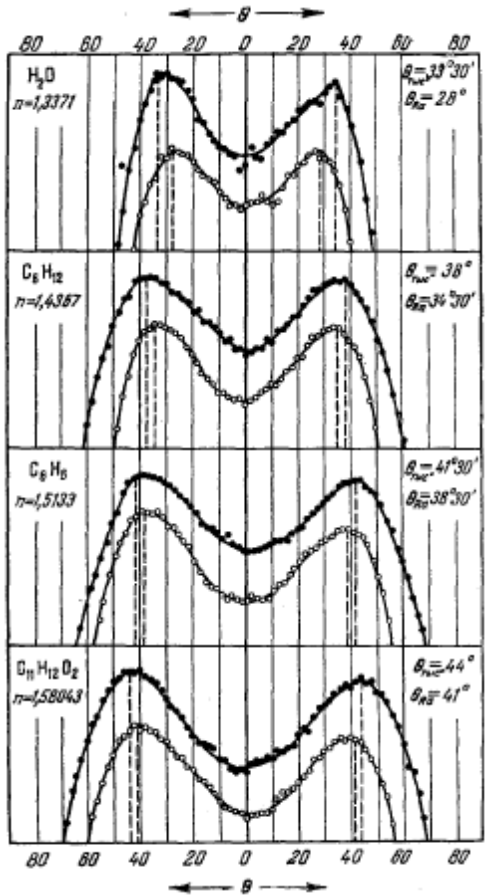

Fig. 1
The observational data of P. A. Cherenkov [31] (see Fig. 4 from [31]) of the angular distribution in angles $0 \leq \theta \leq \theta_m$ for the VCR radiation intensity $I(\theta) > 0$, $I(\theta > \theta_m) = 0$, obtained by passing radium radiation (lower curve) and radiothorium (upper curve) through four liquids having different refractive coefficients. Vertical dotted lines mark the angular position of the interference maximum for $I(\theta)$ at the angle value $\theta = \theta_C < \theta_m$. For example, for water (upper fragment of Fig.1) when irradiated with radium $\theta_C = 28^0$ and $\theta_m = 40^0$.

However, from the same VCR observations [31] shown in Fig.1, it follows that the VCR theory of the Tamm-Frank and Ginzburg, even when taking into account the frequency dispersion of the refractive index, cannot describe the observed position of the maximum cone angle of the VCR and the corresponding threshold for the realization of the VCR effect (see further Tables 1 and [11]-[13]).
In fact, the deviation of the maximum angular position of the VCR cone from the angular position of the VCR interference maximum shown on all curves in Fig.1 is about ten degrees, when the deviation due to the effect of the frequency dispersion of the refractive index is an order of magnitude smaller as for other similar blurring effects is also stated [2].

Indeed, according to Fig.1, the varying of the VCR intensity $I(\theta)$ is observed in a range of angles $0 \leq \theta \leq \theta_m$ (when $I(\theta > \theta_m) = 0$).

For the different angles $\theta$ from this range of angles (for which $I(\theta) \neq 0$) the following dimensionless value of the threshold velocity of an electron corresponds, obtained in the Tamm-Frank theory and in the Ginzburg theory [22]-[25]:

$$\cos\theta = \frac{1}{n\beta};$$
$$\beta = \frac{V_0}{c} = \frac{1}{n\cos\theta} \quad (1)$$

In (1), $c$ - is the speed of light in a vacuum, $n$ - is the refractive index of the medium, which in the VCR theory of the Tamm-Frank and Ginzburg must necessarily exceed one in order to realize the VCR effect; $\theta$ -is the angle between the direction of motion of a charged particle moving at a constant speed $V_0$ and the direction in which the VCR photon is emitted.

However, the theory of VCR [22]-[25] does not describe the entire observed angular distribution of VCR intensity shown in Fig.1. The value of the angle $\theta$ included in (1) is assumed in Tamm-Frank theory [22] to coincide precisely with the value of the angle $\theta = \theta_C$, corresponding to the angular position of the interference maximum observed in the VCR intensity distribution and represented in Fig.1 by dotted lines.

It is generally assumed that the observed and shown in Fig.1, the blurring of the maximum in the intensity distribution of the VCR is mainly due only to the effect of dispersion, manifested by the dependence of the refractive index on the wavelength of light. However, as it follows from the estimates given in the book [2], it can be concluded that the effects of dispersion and scattering alone are probably insufficient for the observed amount $\Delta\theta = \theta_m - \theta_C$ of blurring. This may indicate the need to use a modification of the known condition (1).

For example, it follows from (1) and the first fragment from above in Fig.1 that at the angle $\theta = \theta_C \approx 28^0$, corresponding to the position of the interference maximum of the VCR when irradiated with radium distilled water with a refractive index $n = 1.3371$ for red light, the dimensionless value of the threshold velocity of the charged particle is equal to $\beta = \beta_C \approx 0.847 < 1$.

When taking into account the dispersion, when for violet light the refractive index has a value $n = 1.343$ from (1) for the same value $\beta = \beta_C \approx 0.847$, we obtain the angle value $\theta_D \approx 28.47^0$, which corresponds to a deviation from the angle $\theta_C$ by an amount $\Delta\theta_D = \theta_D - \theta_C \approx 0.47^0$.

This is more than an order of magnitude less than the observed value of the deviation of the maximum angle $\theta_m \approx 42^0$ of the VCR cone from the position of the interference maximum of the VCR intensity, which is equal to $\Delta\theta_M = \theta_m - \theta_0 \approx 14^0$.

Now, on the contrary, we will consider on the basis of (1), with the same value $n = 1.3371$ of the refractive index of water, an estimate of the threshold velocity of a charged particle, when instead of the angle $\theta = \theta_C = 28^0$, we will use, indeed, the threshold value $\theta = \theta_m = 42^0$ for the maximum angle of the cone for the implementation and observation of the VCR effect.

As a result, from (1) we obtain a value $\beta = \beta_m \approx 1.00638 > 1$ exceeding unity of the dimensionless threshold velocity of a charged particle, which is no longer compatible with the concepts of special relativity (SRT).

When the dispersion is taken into account, when the refractive index in water for violet light has a value $n = 1.343$ from (1), it still results in a value $\beta = \beta_m \approx 1.00196 > 1$, that also exceeds one in contradiction with SRT.

Thus, until now, in [1]-[4], [22]-[25], the contradiction with the VCR indicated in [11]-[13] and in the above example is not taken into account, which arises when using condition (1) to obtain an estimate of the threshold for the implementation of the VCR effect by the maximum observed value of the cone angle $\theta_m > \theta_C$.

Indeed, so far, when determining the velocity of charged particles based on data on the refractive index of the medium and the measured value of the cone angle, it is precisely the ratio (1) obtained in the Tamm-Frank theory [1] and in the Ginzburg theory [4], [23] that is used.

For the first time, attention was drawn to this discrepancy between the well-known theory of VCR and the observational data of the VCR realization threshold in [11], where, based on the analysis of all four fragments of Fig.1, Table 1 below was compiled (see Table 1 in [11]):

**Table**

| | $n$ | $n_*$ | $\cos\theta_m^A$ | $\cos\theta_m^B$ | $\beta_*^A = \frac{v_0}{c}$ | $\beta_*^B = \frac{v_0}{c}$ | $\beta^A = \frac{v_0}{c}$ | $\beta^B = \frac{v_0}{c}$ |
|---|---|---|---|---|---|---|---|---|
| $H_2O$ | 1.3371 | 2.2247 | 0.6691 | 0.7431 | 0.6718 | 0.6049 | 1.1177 | 1.0064 |
| $C_6H_{12}$ | 1.4367 | 2.4683 | 0.5 | 0.6428 | 0.8103 | 0.6303 | 1.392 | 1.083 |
| $C_6H_6$ | 1.5133 | 2.6491 | 0.454 | 0.5736 | 0.8315 | 0.6581 | 1.4556 | 1.152 |
| $C_{11}H_{12}O_2$ | 1.5804 | 2.8049 | 0.3584 | 0.5 | 0.9519 | 0.6217 | 1.689 | 1.103 |

The new quantum VCR theory [11]-[13], unlike the quantum VCR theory of Ginzburg (1940) [23], is based on the theory of Abraham [9], [10] and thus allows us to describe the microscopic mechanism of emission of the VCR quantum directly by the medium that is indicated above and in [1]-[4].

The theory of the VCR [11]-[13] explicitly takes into account the change in the energy of the medium when it emits a quantum of the VCR and, as can be seen from Table 1, already well describes the threshold for the realization of the VCR, characterized by the maximum value of the observed angle of solution of the cone of the VCR.

Indeed, in [11], on the basis of a relativistic generalization of the Landau superfluidity criterion [32], instead of (1), the following necessary and sufficient condition for the energetically advantageous realization of an VCR quantum was obtained (by analogy with the mechanism of threshold generation of elementary excitation - a roton in superfluid helium) in the form [2]:

$$\cos\theta \geq \frac{1}{\beta n_*}; n_* = nF(n);$$

$$F = 1 + \frac{\Delta E_m}{E_{ph}} = 1 + \frac{\sqrt{n^2-1}}{n}, n > 1;$$

$$F = \left(1 - \frac{\Delta E_m}{E_{ph}}\right)^{-1} = \frac{1}{1-\sqrt{1-n^2}}, n < 1 \qquad (2)$$

$$0 \leq \theta \leq \theta_m; \theta_m = \arccos\left(\frac{1}{n_*\beta_{th}}\right)$$

$$\beta \geq \beta_{th} = \frac{1}{n_*\cos\theta_m} \equiv \beta_*$$

In (2) $E_{ph}$ -is the energy of the photon emitted by the medium, and $\Delta E_m(n)$ -is the change in the energy of the medium necessary for the possibility of realizing the VCR effect.

The specific determination of the magnitude $\Delta E_m$ of the change in the energy of a medium when it emits VCR photon, which is included in (2), is possible only by using the Abraham representation for the photon momentum in the medium (for more information, see at the end of this paragraph). Note that the threshold for the implementation of the VCR in (2) corresponds to the condition of an energetically favorable generation of the VCR quantum by the medium. Condition (2) was obtained in [11] with a relativistic generalization of the Landau criterion for energetically favorable vortex generation [32].
That is why in (2) there is an inequality sign, the use of which, as in the Landau theory [32], follows due to the inevitable presence of dissipative factors leading to a violation of the strict energy balance during the birth of a quantum. On the contrary, when there are no dissipative factors, then, in accordance with the law of conservation of energy in the quantum VCR theory, also considered in [11], (2) should be considered a sign of strict equality, as in condition (1) obtained in the theory of quantum mechanics by Tamm-Frank and Ginzburg [22], [24].

---

[2] The results of the paper [11] were repeatedly discussed with B. M.Bolotovsky before its publication. According to Boris Mikhailovich, V.L. Ginzburg also showed interest in this publication.

However, due to the inevitable presence of dissipation, which leads to a violation of the conservativeness of the system in (2), the sign of inequality that should be considered, which corresponds to the observational data [31] (see Fig.1). Indeed, according to [31], the VCR is observed in the entire range of angles $0 \leq \theta \leq \theta_m$, satisfying precisely the strict inequality (2), in contrast to the equality in (1) of the well-known Tamm-Frank and Ginzburg VCR theory (see Table 1).

Table 1 shows the threshold values of the parameter $\beta_*$ defined in (2) and corresponding to the observational data [31] presented in all four fragments of Fig.1. Table 1 shows that all the threshold values of this parameter do not exceed one, unlike the threshold values of the parameter $\beta$ following from the application of the ratio (1) the well-known theory of VCR [1]-[4], [22]-[25].

Indeed, it follows from the data of the Cherenkov experiment [31] that a sufficient condition for the implementation of the VCR observed in the range of angles $0 \leq \theta \leq \theta_m$, according to Fig.1, is determined only from the inequality $\beta = V/c > 1/n\cos\theta_m$ following from (1) or from the inequality $\beta = V/c > \beta_* \equiv 1/n_*\cos\theta_m$ corresponding to (2).

In contrast to condition (1) the, condition (2) is valid not only for the refractive index $n > 1$ of an isotropic medium, but also for the values $n < 1$ that determine in (2) the threshold for the emission of transverse VCR waves in isotropic plasma.

Note that, unlike the well-known necessary condition for the implementation of the VCR, which implies the requirement that the particle velocity exceed the value of the phase velocity of light in the medium, another condition necessary for the implementation of the VCR follows from (2):

$$\beta \geq \frac{1}{n_*} \qquad (3)$$

Thus, in the new quantum VCR theory [11]-[13], there is no restriction $n > 1$ on the value of the refractive index, which is known specifically for the macroscopic VCR theory of Tamm-Frank and for the quantum VCR theory of Ginzburg, since these theories do not explicitly take into account the VCR mechanism of photon emission by the medium.

This makes it possible to use the theory of microwave radiation [11]-[13] not only for a more accurate interpretation of the VCR observational data in Cherenkov detectors, but also for the analysis of the observed effects of X-ray radiation in isotropic plasma [33]-[37].

However, the purpose of this article is not only to draw attention to the possibility of expanding the scope of the new quantum VCR theory [11]-[13], based on the well-known theory of Abraham [3], [9], [10], confirmed by experiment [6]-[8], and describing the electromagnetic field (EMF) in a media.

An equally important task facing this publication is to eliminate the still existing reason that prevents the use of Abraham's theory in describing the EMF in the medium instead of the continued widespread use, on the contrary, of Minkowski theory.

This reason is the still popular (doubtful, as shown above when comparing (1) and (2) in Table 1) idea of the success of the application in the well-known quantum VCR theory [23]-[25] of the Minkowski theory [5], which contradicts the theory of Abraham [9], [10] when describing the EMF in the medium.

Due to this view, outlined in [38], the Minkowski theory [5] continues to be widely used even after it was rejected back in 1975 by experiments that established the existence of the Abraham force [6]-[8].

As a result, the so-called dilemma of the Minkowski and Abraham theories is still being discussed, and it is Minkowski theory that is widely used, as well as the well-known quantum VCR theory based on it, leading to the condition for the implementation of the VCR effect in the form of relation (1).

However, it was noted above that the idea of the success of the quantum VCR theory [23]-[25] clearly contradicts experimental data on the threshold for the implementation of VCR due to the impossibility of describing the microscopic mechanism of VCR emission directly by the medium.

Thus, the parity attitude towards the theories of Minkowski and Abraham that still persists is largely due to the lack of a well-known alternative quantum VCR theory based on Abraham's theory.

This gap has been eliminated in the new quantum VCR theory [11]-[13], based on Abraham's theory, and, as shown in Table 1, it leads to a better correspondence with observational data compared with the quantum theory of quantum mechanics based on Minkowski theory.

Only in Abraham's theory, as shown below, it is possible to explicitly take into account the change in the energy $\Delta E_m$ of the medium when it emits a photon according to the well-known microscopic mechanism of the VCR effect [1]-[4].

In fact, in the new quantum theory of electromagnetic radiation [11]-[13], when writing the energy and momentum balance equations describing the effect of photon emission by an electromagnetic radiation medium, the following representation is used for the photon momentum in the medium, corresponding to the Abraham theory [38]:

$$\vec{p}^A(n>1) = \frac{E_{ph}}{cn}\frac{\vec{k}}{|\vec{k}|}; n>1; \vec{p}^A(n<1) = \frac{E_{ph}n}{c}\frac{\vec{k}}{|\vec{k}|}; n<1 \qquad (4)$$

In (4) $\vec{k}$ -is a wave vector characterizing the direction of propagation of a photon when it is emitted by a medium at an angle $\theta$ - relative to the direction of the charged particle velocity $\vec{V}_0$. In (1) and (2), the magnitude of the modulus $V_0 \equiv |\vec{V}_0|$ of this velocity is used.

The very possibility of considering the medium in [11]-[13] as a photon emitter in VCR is due to the fact that, unlike Minkowski theory, Abraham's theory allows for the introduction of a real rest mass of a photon in the medium for a medium with a refractive index $n \neq 1$, using the well-known representation for the mass $m_{ph}(n \neq 1) > 0; m_{ph}(n = 1) = 0$ of any relativistic particle through its energy and momentum [38]:

$$m^2 c^4 = E^2 - p^2 c^2 \qquad (5)$$

The finiteness of the mass of a photon in a medium is due to the fact that, unlike a vacuum, a photon in a medium is not free, but is a quasi-particle, named in the article by I. M. Frank [39] (see also [40]) a quasi-photon.

When substituting representations (4) for the momentum of a photon in a medium into equation (5), we obtain the following representation for the mass of a photon in any medium with a refractive index other than unity:

$$m_{ph} = \frac{E_{ph}\sqrt{n^2-1}}{c^2 n}, n > 1; m_{ph} = \frac{E_{ph}}{c^2}\sqrt{1-n^2}, n < 1 \qquad (6)$$

To obtain the threshold condition for the VCR effect in the form of (2), assume that when a medium emits a photon with energy $E_{ph}$, the energy of the medium decreases by an amount related to the mass of the photon (6) and the ratio takes place:

$$\Delta E_m = m_{ph} c^2 \qquad (7)$$

The threshold condition (2) for the emission of the VCR photon by a medium was obtained in [11]-[13] (see also section 6 below) from the momentum and energy balance equations precisely taking into account relation (7), when the representation for the actual positive mass of a photon in the medium in the form (6) is acceptable.

Within the framework of Minkowski theory [5], [38], it is no longer possible to determine the positive and real mass of a photon in a medium [38], [39].

In fact, according to Minkowski theory, the momentum of a photon in a medium is defined in the following form, usually used in the quantum theory of the Ginzburg quantum field [23]-[25]:

$$\vec{p}^M(n \neq 1) = \frac{E_{ph} n}{c} \frac{\vec{k}}{|\vec{k}|}; \forall n \neq 1 \qquad (8)$$

It is not difficult to verify that what is usually considered in the well-known quantum theory is [23]-[25], [38] in the case $n > 1$ of (8) and (5), only a purely imaginary value of the photon mass in the medium is obtained:

$$m_{ph}^M = i \frac{E_{ph}}{c^2 n}\sqrt{n^2-1} \qquad (9)$$

Relation (9) is given in [38] (see page 142 in [38]), it is also obtained by substituting the momentum of the photon (8) into the relativistic relationship (5) between the mass of any particle and its momentum and energy.

It is obvious that the complex value of the mass of a photon (9) in a medium, following from the Minkowski theory and representation (8), cannot be taken into account in the equations of energy and momentum balance describing the VCR effect. Apparently, it is in this connection, as well as because of the negative mass of a photon in a medium obtained in [39] on the basis of the Minkowski representation (8), that [38] concludes that it is inappropriate to use the concept of photon mass in a medium.

Therefore, it is more natural to consider the impracticability of using the mass (9) of a photon in a medium, corresponding to the Minkowski representation (8) for the photon momentum in a medium with the refractive index $n > 1$ of an isotropic medium.

On the contrary, in the new quantum theory of photons [11]-[13], the expediency of using the concept of the mass of a photon in a medium (6) is shown, but only on the basis of using Abraham's theory and representation (4) for the momentum of a photon in a medium (in section 5).

On the basis of Abraham's theory, a new interpretation of some well-known experiments is also given (in the next section 3), interpreted earlier on the basis of Minkowski's theory without taking into account Abraham's force.

### 3. The dilemma Abraham's and Minkowski's theories and VCR theory

An experiment sometimes allows us to reject one of two contradictory theories that provide an equally valid mathematical description of the observed phenomena and processes.

However, this opportunity was missed about half a century ago when solving the still-discussed problem of choosing between two contradictory theories of EMF in the environment, proposed more than a hundred years ago by Minkovsky [5] and Abraham [9], [10].

It is important that only the theory of Abraham [9],[10] predicts the existence of an additional, measurable volumetric force acting on any medium (including an electrically neutral, stationary, homogeneous and isotropic medium) from the side of the EMF [38], [41]. This so-called Abraham force is absent from Minkowski theory. [5], [38], [41].

The fundamental initial difference in representations for the EMF momentum density in the medium in the theories of Minkowski $\vec{g}^M = n^2 \vec{g}^A$ и and Abraham $\vec{g}^A = \vec{S}/c^2$ is uniquely related precisely to the presence of the Abraham force density $\vec{f}^A = (\varepsilon\mu - 1)c^{-2}\partial\vec{S}/\partial t$ only in Abraham's theory. Here $\vec{S} = c[\vec{E} \times \vec{H}]/4\pi$ - is the vector of the EMF energy flux, $\varepsilon\mu = n^2$, where $n \neq 1 (n > 1; n < 1)$- is the refractive index of the isotropic medium, $c$ -is the speed of light in vacuum [38], [41].

Thus, the experimental detection of any example of a finite Abraham force is a necessary and sufficient condition for a rigorous mathematical proof of the validity of the Abraham representation and the inadmissibility of an alternative Minkowski representation when describing an EMF momentum in a medium.

Moreover, the formulation of the problem of measuring the Abraham force makes it unnecessary to directly measure the EMF pulse or photon momentum in the medium. Indeed, such measurements may obviously allow ambiguous explanations for the observed effect of the interaction of electromagnetic radiation and the environment, which clearly makes it difficult to choose between alternative theories of Minkowski and Abraham.

That is why Pauli, already about a hundred years ago, noted that the choice between Minkowski theory and Abraham's theory for EMF in the medium could be made only by detecting the Abraham force [41]. In this regard, in his book [41] (see page 152 in [41]) it is written: "The Abraham energy-momentum tensor leads to the addition of a term to the ponderomotor force in resting bodies. Due to the smallness of this term, it is unlikely to be possible to assume a practically feasible experimentum crucis for comparing the theories of Minkowski and Abraham."

Nevertheless, just such a decisive (experimentum cruces) experimental measurement of the Abraham force [6]-[8] was nevertheless carried out, albeit several decades after the discussion of the dilemma of the theories of Minkowski and Abraham began.

This was first done in 1968 by James [6] and in 1975 by Walker and his co-authors [7], [8] an even more direct and fairly accurate (about 5%) measurement of the Abraham force predicted in the Abraham theory was obtained.

As a result, the validity of Abraham's theory was unequivocally recognized not only by supporters of Abraham's theory, which, in addition to Pauli and Einstein, includes Landau [3], but also by well-known supporters of Minkowski's theory [38],[42], [43]. In particular, in the book by Landau and Lifshitz [3], the Abraham representation is already used for the EMF momentum in the medium, which is uniquely interrelated with the Abraham force density (see (56.16) and (56.18) in [3]).

However, instead of logically acknowledging the "fallacy" of Minkowski's alternative theory, in [38], [42], [43] it is noted that such recognition is possible only with an "excessively formal approach to the case" [38].

Moreover, it was considered not only possible, but even useful and successful to apply the Minkowski representation in the quantum theory of Vavilov-Cherenkov radiation (IVH) [38],[42], [43].

In turn, the success of the quantum theory of the Ginzburg quantum field (1940) [23], which has not been doubted until now (noted above and in [11]-[13]), is based only on a good agreement of the conclusions of this theory with the condition (1) of the realization of the effect of the field, obtained previously in the macroscopic theory of the Tamm-Frank IC [22].

As a result of this idea of "usefulness and success", it is Minkowski's theory, and not Abraham's theory, in direct contradiction with the above conclusion of experiments [6]-[8], that is still widely used in describing the EMF in medium [44]. Subsequent experimental measurements of the Abraham force in [45], [46] did not affect this in any way, as did the creation of the quantum VCR theory [11]-[13], based on the theory of Abraham.

Indeed, in reputable publications [38],[42], [43] for the first time, the concept of usefulness, although formally "incorrect" Minkowski theory appeared.

This is best illustrated by the following set of quotations from V. L. Ginzburg's wonderful book [38], taken from pages 320-321 [38], where, after discussing the ambiguity in the choice of expression for the EMF energy-momentum tensor, it is written:

"Force density is a completely different matter, which is, at least in principle, an unambiguous and measurable quantity. In this regard, the fate of the "dispute" about the Abraham and Minkowski tensors is ultimately decided as a result of the choice of an expression for the force.

There is no doubt about the reality of this force, although it has only recently been measured.

Thus, the issue is unequivocally resolved "in favor " of the Abraham tensor. The objections to the choice of this tensor found in the literature are not substantiated.

All of this makes it possible to consider the Abraham tensor "correct", but, as it seems to us, it is possible to declare the Minkowski tensor "incorrect" only by approaching the problem somewhat formally. In fact, in most situations, the results obtained using the Abraham and Minkowski tensors are completely identical. This makes it possible, in appropriate cases, not only to use the Minkowski tensor, but even to consider its use quite appropriate if some kind of simplification is achieved thereby. Therefore, the Minkowski tensor should hardly be declared "erroneous"; rather, it represents some auxiliary concept that may well be used. This does not in any way damage the "prestige" of the more fundamental and, if you will, "true" energy-momentum tensor of the electromagnetic field in the medium. An analysis of the laws of conservation of energy and momentum of electromagnetic waves (photons) in the medium confirms and illustrates the remark just made."

In [11]-[13] and the present work, examples are given that refute the generality of the statement made in [38] about the identity of the results obtained based on the application of the theories of Minkowski and Abraham. In fact, these theories give identical conclusions, coinciding with each other, only when considering EMF in a vacuum, for which the refractive index $n=1$ is unity.

Even with small deviations from unity in the limit $n-1 \ll 1$, one would expect a small deviation in the conclusions of the theories of the IHR based on the theories of Minkowski and Abraham. However, this is not the case, and the differences can be many orders of magnitude in this limit [13] (see also Fig.3 below).

We also show that the argumentation given in [38], which substantiates the claim of the identity of the conclusions of the theories of Minkowski and Abraham, cannot serve as a basis for the idea about the convenience and usefulness of Minkowski's theory.

Indeed, [38] analyzes the law of conservation of momentum used in the quantum VCR theory of the Ginzburg [23], and on the basis of which the statement given in [38] is substantiated (see also [42], [43]) about the identity of the results of using the theories of Minkowski and Abraham.

In this regard, the following relation is considered, which is considered universal (see (44) in [42] and (13.22) in [38]):
:

$$\vec{p}^M = \vec{p}^A + \vec{F}^A \qquad (10)$$

The left-hand side of equation (10) uses the representation for the photon momentum in Minkowski theory discussed in (8) [38]:

$$\vec{p}^M = \frac{1}{T}\int dt \int d^3x \vec{g}^M = \frac{E_{ph} n}{c} \frac{\vec{k}}{k}; \forall n \neq 1, \qquad (11)$$

Similarly, in the right part of (10), the representation [38] is used for the momentum of a photon in a medium, according to the theory of Abraham, presented above in (4):

$$\vec{p}^A = \frac{1}{T}\int dt \int d^3x \vec{g}^A = \frac{E_{ph}}{nc} \frac{\vec{k}}{k}, n > 1 \qquad (12)$$

$$\vec{p}^A = \frac{1}{T}\int dt \int d^3x \vec{g}^A = \frac{nE_{ph}}{c} \frac{\vec{k}}{k}, n < 1 \qquad (13)$$

In (11)-(13), averaging over the period of high-frequency EMF oscillations with a period $T$ is assumed.

The pulse that is transmitted to the medium during the emission of EMF waves is determined by the magnitude of the space-integral Abraham force pulse according to the estimate given in [38] (see (13.21) in [38]) and having the following values at $n > 1$ (see also [26] and (B.5) in Appendix B):

$$\vec{F}^A(n > 1) = \int dt \int d^3x \vec{f}^A = (n^2 - 1)\vec{p}^A = \frac{(n^2-1)}{cn} E_{ph} \frac{\vec{k}}{k}, n > 1 \qquad (14)$$

Note that (14) also shows the time-average magnitude of the force pulse, which acts only as long as the wave or set of waves enters the medium or is emitted by the source [38] (see the footnote on page 322 in [38]). At the same time, unlike (11)-(13), the time integral in (14) no longer means averaging over the period of high-frequency EMF oscillations in the medium, but corresponds to the replacement $\int dt \partial \vec{g}^A / \partial t \to T^{-1}\int dt \vec{g}^A \equiv \langle \vec{g}^A \rangle$, where the angle brackets indicate averaging over high-frequency EMF oscillations used in (11)-(13). We also note the relation $\langle \vec{f}^A \rangle = (n^2 - 1)\langle \partial \vec{g}^A / \partial t \rangle = 0$, which is used in [43] as an argument about the insignificance of taking into account the Abraham force when considering high-frequency phenomena in the environment. However, this argument contradicts the relation (10) given in [38], in which taking into account the finite magnitude of the momentum of the Abraham force is the determining meaning of this relation.

Thus, the "convenience" of using Minkowski theory mentioned in [38] consists in using the representation of the photon pulse in a medium in the Minkowski form (8), (11), instead of the right-hand side of equation (10) or the sum of representations (12) and (14) of Abraham's theory given in [38] (see (13.22) in [38]):

$$\vec{p}^M = \vec{P}^A(n>1) \equiv \vec{p}^A(n>1) + \vec{F}^A(n>1) = \frac{E_{ph} n}{c} \frac{\vec{k}}{k} \qquad (15)$$

However, the representation (14) is defined only for values $n>1$, and for the refractive index $n<1$, instead of (14), the Abraham force impulse should be used in the form (see [26] and conclusion (B.5) in Appendix B):

$$\vec{F}^A(n<1) = (n^2-1)\vec{p}^A = \frac{(n^2-1)n}{c} E_{ph} \frac{\vec{k}}{k}, n<1 \qquad (16)$$

As a result, taking into account (16), it is the case $n<1$ that gives an example of a violation of the universality of relation (10) claimed in [38], when the Minkowski representation on the left side of relation (10) is no longer equal to the sum of the two terms on the right side of (10).

Indeed, in the case of the sum of the momentum of a photon in a medium in the form of Abraham (13) and the momentum of the Abraham force (16) no longer leads to a momentum of a photon in the medium in the form corresponding to Minkowski theory and defined in (8) and (11). In this case, the sum of impulses transmitted to the medium, according to Abraham's theory, already has a completely different form and is described by the equality:

$$\vec{p}^M > \vec{P}^A(n<1) \equiv \vec{p}^A(n<11) + \vec{F}^A(n<11) = \frac{n^3}{c} E_{ph} \frac{\vec{k}}{k} \qquad (17)$$

Thus, for the case $n<1$ it is stated in (17) the invalidation of the notion of the "convenience" of using Minkowski theory, introduced in [42], [43], [38] based on the relation (15). There are also indirect manifestations of the Abraham force impulse in many experiments describing the interaction of electromagnetic radiation and the medium [47]-[53]. This can be seen, as shown in the next section, if we take into account the Abraham force and its momentum when analyzing experimental data.

It still seems important, on the basis of equality (15), to understand the reason for the formal correspondence between the result of the quantum VCR theory [23] and the macroscopic theory of the Tamm–Frank microwave oven [22].

Indeed, in the quantum theory [23], [38], the following representation is obtained (where $E_{ph} = \hbar\omega$ см. (7.3) в [38]):

$$\cos\theta_0 = \frac{c}{V_0 n(\omega)}\left(1 + \frac{\hbar\omega(n^2-1)}{2mc^2\Gamma_0}\right);$$

$$\Gamma_0 = \frac{1}{\sqrt{1-V_0^2/c^2}}$$

(18)

In (18) $m$ - is electron's mass. For the optical wavelength range of the VCR corresponding to the limit $E_{ph}/mc^2 \ll 1$, the second term in the bracket can be ignored. In this case, the condition for the implementation of the VCR (18) exactly coincides with the threshold of the VCR obtained in the macroscopic theory of Tamm-Frank (1937) [22] and presented in condition (1).

The formal fulfillment of equality (15) for the case $n > 1$ allows us to explicitly determine the possible range of applicability of the Abraham force momentum (14) in the conservation law the pulse.

Indeed, the Tamm-Frank macroscopic theory of VCR [22] describes a coherent VCR field that already exists in the medium. Therefore, from the correspondence of the Ginzburg theory [23] with the theory [22], it follows that equality (15) should also be valid precisely for describing the already realized effect of VCR in the medium, and not for the threshold of realization of this effect. Therefore, taking into account the Abraham force pulse (14) is no longer necessary to determine the threshold for the emission of the VCR by the medium, described by condition (2) in the new VCR theory [11]-[13].

The ratio (10) or (15) was also used in [54] to obtain a new explanation of the experimental results [55], [56] based on Abraham's theory, although previously, before the publication [54], these results were interpreted only in favor of Minkowski's theory.

In this regard, it is desirable to have verification not only for values $n > 1$, as in experiments [55], [56], but also for cases $n < 1$ in the experiments [57]-[59] which are also interpreted before in support of the Minkowski theory [60]-[63].

In the next section, the results of the experiment [59] are analyzed, which examines the interaction of a photon with a single condensate atom for EMF wave frequencies close to resonance, when it is necessary to take into account the case of the high-frequency part near resonance at refractive index values.

Figure 2 demonstrates that the application of Minkowski theory leads to the lack of the possibility of interpreting experimental data [59] in the field of parameters corresponding to the case $n < 1$. However, when using of the Abraham force impulse in the form (16), such an interpretation is given based on the use of an estimate of the resulting impulse in the form (17) transmitted from the EMF to the condensate atom.

## 4. The Abraham force and the interaction of EMF with dielectrics

Let us now consider in more detail the mathematical basis for the contradiction between the theories of Abraham and Minkowski, which is given in a relatively simple form in chapter 13 of V. L. Ginzburg's book [38] and briefly described later in this section.

We show that it is the Abraham force, which is additional to the Lorentz force in describing the interaction of EMF and matter, that is the key factor determining the fundamental difference between Minkowski and Abraham's theories and the possibility of solving the dilemma of these theories based on experiments. [6]-[8], [45], [46] by direct measurement of this force.

We will also give a specific example that shows the need to take into account the Abraham force in order to interpret experimental data [59], which turns out to be impossible without taking this force into account in the framework of Minkowski theory, which does not allow an explanation of part of the observational data presented in Fig.2.

### 4.1 The mathematical basis of the Minkowski and Abraham theories dilemma

The classical Maxwell equations for EMF in a stationary, non-magnetic, non-absorbing, and transparent medium without dispersion have the form ($\vec{B} = \vec{H}; \mu = 1; \varepsilon = n^2 = const$):

$$rot\vec{H} = \frac{4\pi}{c}\vec{j}_{ext} + \frac{n^2}{c}\frac{\partial \vec{E}}{\partial t} \tag{19}$$

$$rot\vec{E} = -\frac{1}{c}\frac{\partial \vec{H}}{\partial t} \tag{20}$$

$$n^2 div\vec{E} = 4\pi\rho_{ext} \tag{21}$$

$$div\vec{H} = 0 \tag{22}$$

Based on the use of equations (19)-(22) in [38], the following conservation equation for the EMF momentum density $\vec{g}^A$ is derived, written in the Abraham form (see (13.8) in [38]):

$$\frac{\partial g_\alpha^A}{\partial t} - \frac{\partial \sigma_{\alpha\beta}}{\partial x_\beta} = -f_\alpha^L - f_\alpha^A; \alpha = 1,2,3 \tag{23}$$

$$g_\alpha^A = \frac{1}{4\pi c}\varepsilon_{\alpha\beta\gamma}E_\beta H_\gamma \tag{24}$$

$$\sigma_{\alpha\beta} = \frac{1}{4\pi}\left[n^2 E_\alpha E_\beta + H_\alpha H_\beta - \frac{1}{2}w\delta_{\alpha\beta}\right];$$

$$w = \frac{1}{8\pi}\left(n^2 E^2 + H^2\right) \tag{25}$$

$$f_\alpha^L = \rho_{ext} E_\alpha + \frac{1}{c}\varepsilon_{\alpha\beta\gamma} j_\beta^{ext} H_\gamma \qquad (26)$$

$$f_\alpha^A = \frac{(n^2-1)}{4\pi c}\varepsilon_{\alpha\beta\gamma}\partial(E_\beta H_\gamma)/\partial t \qquad (27)$$

In the case of a weak frequency dispersion of the refractive index $n$ of the medium, a replacement $n^2 \to d(\omega n^2(\omega))/d\omega$ [64] should be made in the representation for the EMF energy density $w$ in (25). In equation (23), the Abraham force density defined in (27) has the same form for any refractive index values $n \ne 1: n > 1; n < 1$ other than unity.

In representations (24)-(27), $\varepsilon_{\alpha\beta\gamma}$ - is an absolutely antisymmetric unit pseudotensor of the third rank, where $\varepsilon_{123} = 1$ and $\sigma_{\alpha\beta}$ - is the Maxwell stress tensor.

The right-hand side of the EMF momentum conservation equation (23), in addition to the Abraham force (27), also includes the Lorentz force (26), which also describes the effect of EMF on the medium, but only if there are free charges and currents in it.

The minus sign on the right side of equation (23), written to determine the balance of forces and momentum applied to EMF, is due to the fact that the sum of forces (26) and (27), taken with a plus sign, represents the force acting on the medium, according to Newton's third law [38].

In Minkowski theory, the conservation equation for the EMF pulse density in a medium differs in form from equation (23) not only due to the absence of the Abraham force density (27) on the right side of the equation, but also from the definition of the EMF pulse density in a medium other than in (24) (see further (29)).

Therefore, in Minkowski theory, the balance equation for the EMF pulse in a medium has the form [38]:

$$\frac{\partial g_\alpha^M}{\partial t} - \frac{\partial \sigma_{\alpha\beta}}{\partial x_\beta} = -f_\alpha^L; \alpha = 1,2,3 \qquad (28)$$

$$g_\alpha^M = \frac{n^2}{4\pi c}\varepsilon_{\alpha\beta\gamma} E_\beta H_\gamma \qquad (29)$$

Only an experimentally obtained confirmation in [6]-[8], [45], [46] the reality of the existence of an Abraham force of the form (27) allows us to make an unambiguous choice between equations (23) and (28) in favor of representing Abraham for the law of conservation of EMF momentum in the form (23). The possibility of such a choice predicted by Pauli [41] was indeed noted in [42], [43] and [38] after the first experiments [6]-[8].

Indeed, as indicated in [38], equation (23) for the EMF momentum in the Abraham form (24) is mathematically strictly equivalent to equation (28) for the EMF momentum in the Minkowski form (29). This can be confirmed by transferring the Abraham force (27) from the right side of equation (23) to its left side, which leads exactly to equation (28), but already when using the EMF pulse in the Minkowski representation medium (29). However, this formal mathematical correspondence does not imply the equality of Minkowski's physical theory based on (28), (29) and Abraham's theory using the representation of the law of conservation of EMF momentum in the form of (23), (24).

Thus, for several decades before 1975, there was still a formal mathematical equality of representations of the theories of Abraham and Minkowski for the law of conservation of the EMF momentum and the EMF momentum itself in the medium. Therefore, the statement about such equality actually had the character of a strict mathematical theorem until 1975, according to which there is no possibility of choosing between the physical theories of Minkowski and Abraham when describing EMF in any media.

Therefore, the "experimentum crucis", noted by Pauli in [41], and implemented in [6]-[8], became a counter example refuting the indicated "mathematical theorem" on the equality of the theories of Minkowski and Abraham.

Later, in experiments [45], [46], data consistent with the theoretical value (27) were also obtained for the value of the Abraham force with an accuracy of <1%, exceeding the accuracy of the first experiments [6]-[8].

Thus, to reject the equality of the theories of Abraham and Minkowski and the corresponding logical recognition of the fallacy of Minkowski's theory, is possible due to the experiments [6]-[8], [45], [46].

Nevertheless, as noted above, there was no such unequivocal recognition of the fallacy of Minkowski's theory, and this, as a result, served as the basis for doubts about the applicability of Abraham's theory to describe high-frequency EMFs in the medium [44], [60]-[62], including a description of the VCR effect.

For example, in [43], [60]-[62] it is noted that since the experiments [6]-[8] were conducted only for low-frequency EMFs, they do not provide direct evidence to make a choice between the representations of Abraham and Minkowski for the EMF pulse in the medium.

However, even if we accept the well-known idea [42], [62] about the extremely small value of the time averaging result for the Abraham force in the case of high-frequency EMFs, this does not negate the very fact of the existence of the Abraham force, which the authors have no doubt [42], [43], [38] and [60]-[62]. This, in turn, makes it possible to unequivocally recognize the fallacy of Minkowski's theory, which asserts the validity of the EMF momentum conservation equation in the form (28), which lacks the Abraham force for any EMF frequencies. In this case, the representation for the EMF momentum in the medium in the Minkowski form (29), which is included in equation (28), automatically becomes equally unacceptable.

Indeed, the relatively recent direct measurement of the Abraham force in [45], [46] gave the authors of these works the basis for an unambiguous conclusion about the refutation of the Minkowski theory, leading to the momentum balance equation (28), in which this force is absent. In the experiments to detect the Abraham force in [45], [46], as in earlier experiments [6]-[8], the Abraham force arises from a crossed oscillating electric field and a static magnetic field, but no longer in a solid dielectric, as in [6]-[8], but in a gaseous medium.

Thus, any particular fact of the existence of the Abraham force, which has been experimentally discovered, clearly implies a violation of the mathematical equality of the formulations of Abraham (23), (24) and Minkowski (28), (29) for the law of conservation of EMF momentum in a medium. This also corresponds to the requirement of unambiguous recognition of the error of the Minkowski representation (29) for the EMF pulse in the medium and the subsequent representation for the photon momentum in the medium (11) and (8), used in the well-known quantum VCR theory [23]-[25] in obtaining the relations (18) and (1).

It is important to note that the refractive index of the medium in equation (23) can either exceed one or be less than one, since in both cases the value of the Abraham force density (27) is nonzero.

The previous section shows that the concept of "convenience" of using Minkowski theory, introduced in [38], is unjustified, based on the relation (10) established in [38], applicable only for the case $n > 1$ in the form (15).

Thus, taking into account the resolution of the dilemma of the theories of Abraham and Minkowski, which actually took place in 1975, it would have been possible, long before the works [11]-[13], back in [42], [43],[38], instead of substantiating the "usefulness" of Minkowski's theory, raise the question of the need to create a the basis of Abraham's theory is precisely the microscopic quantum VCR theory of the VCR emission by the medium.

However, this did not happen, and, conversely, over time, the concept of a "useful" Minkowski theory was transformed into the idea of a "correct" Minkowski theory [44]. Indeed, in many works (see [65]-[71]), the belief in the correctness of Minkowski theory and its modifications, rather than Abraham's theory, clearly dominates.

This also provided the basis for surprising in [72] regarding the representation of the EMF momentum in the medium.

The following section provides an example of an analysis of experimental data from the work of Campbell et al. [59], in which, as it turned out, a complete interpretation of these data is possible only on the basis of Abraham's theory.

On the contrary, until now, it was the experimental data [59] that were interpreted solely in favor of the Minkowski theory [60]-[63]. Moreover, the considered example shows that the declared in [42], [43],[38] the "convenience" of using Minkowski's theory even hinders the relevant analysis of experimental data [59] based on Abraham's theory.

### 4.2 The Abraham force and the observed interaction of a photon with a BEC

In [59], for the first time, observations of a systematic change in photon recoil pulses, depending on the refractive index $n; n > 1, n < 1; |n-1| \ll 1$ of a rarefied gas consisting of atoms $^{87}Rb$, are presented. The recoil frequency associated with the recoil energy $E_{rec}(n) = \hbar\omega = p_{rec}^2(n)/2m$, is determined using a two-pulse interferometer with a light array when using laser radiation with a frequency and wavelength $\lambda = 780nm$ close to resonance.

In [59], based on the measurement data presented further in Fig.2 (see also Fig.3 in [59]), it was concluded that the experimentally observed dependence of the recoil energy on the refractive index should correspond to the resulting momentum value $p_{rec} = p^M = E_{ph}n/c$, transmitted to the atom at the refractive index value $n > 1$.

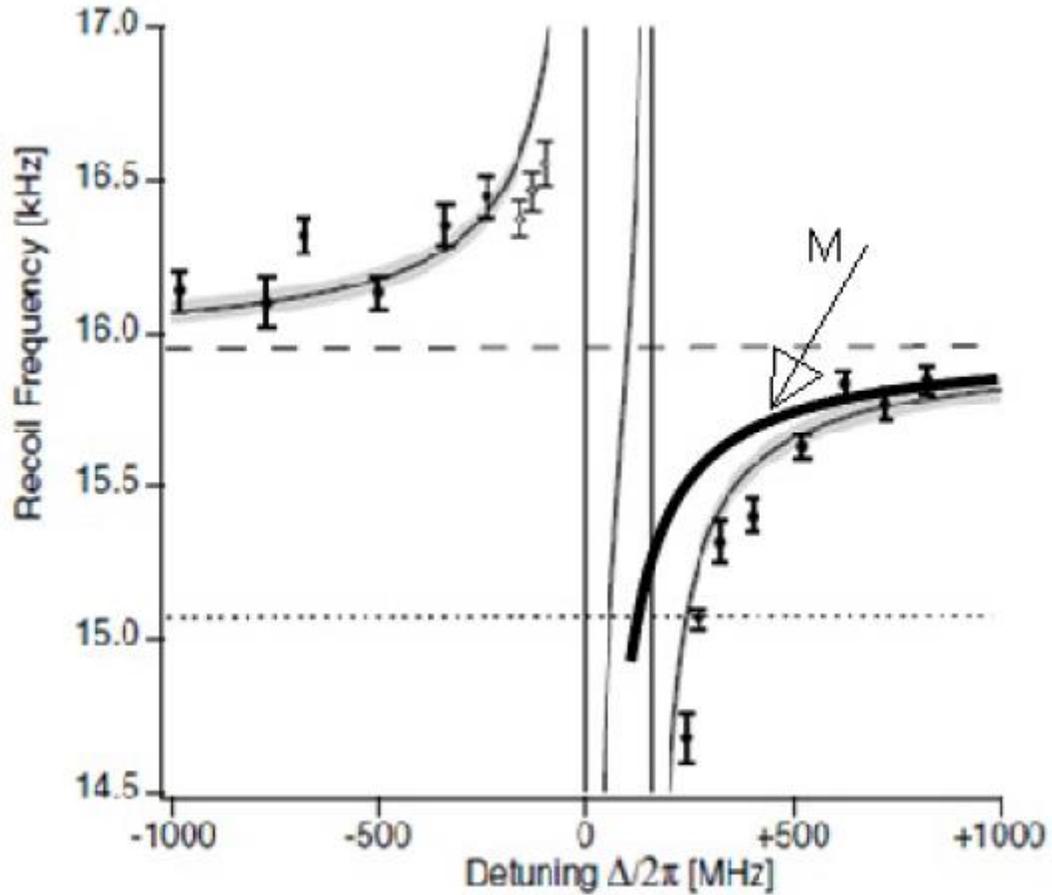

Fig.2
The recoil frequency $\omega(n(\Delta/2\pi))$ as a function of the deviation $\Delta/2\pi$; shows the dispersion effect of the refractive index $n$: at values $n > 1$ for a negative value $\Delta/2\pi < 0$ and $n < 1$ at a positive deviation $\Delta/2\pi > 0$ from the resonant frequency.

The average density of the condensate is $1.14(4) \times 10^{14} cm^{-3}$, which leads to a shift of the average field by $\rho U/\hbar = 880$ Hz. The shaded area shows the expected recoil frequency $\omega = 4n^2 \omega_{rec} + \rho U/\hbar$. The dotted line is on the line corresponding to the expected value $\omega = 4\omega_{rec} + \rho U/\hbar$ without taking into account the effect of refraction. The horizontal straight line, indicated by dots, is at the frequency level $4\omega_{rec} = 15\,068$ Hz, corresponding to the two-photon vacuum recoil. Fig. 2 corresponds to Fig.3 from [59]. The arrow with the letter M points to a solid thick curve, which corresponds to the use of Minkowski theory to interpret experimental data corresponding to $n<1$ at the positive value $\Delta/2\pi > 0$.

The relation $p_{rec} = p^M = E_{ph} n/c$ coincides with the Minkowski representation (8) (or (11)) for the momentum of a photon in a medium, which in [60], as noted above, is interpreted precisely as an indicator of the usefulness of Minkowski theory.

However, the conclusion obtained in [59] can also be explained based on the use of the Abraham representation for the photon momentum in a medium, if the values take into account the ratio (10) given in [38] or the equivalent equation (15) for the refractive index $n>1$. Indeed, according to (10) or (15), it is the equality $p_{rec} = P^A(n>1) = E_{ph} n/c$ that holds.

Indeed, the value of the refractive index $n>1$ in Fig. 2 corresponds to the upper left relatively low-frequency branch of the dispersion curve, which corresponds well to the resulting momentum (15) transmitted from the EMF atom when it absorbs a photon.

In addition, [59] noted the lack of explanation for the observed asymmetry in estimates of the characteristic attenuation times $\tau_C$ of the interference fringes between the red (low-frequency, where $n>1$ and $\tau_C = 347(20)\mu s$) and blue (high-frequency, where $n<1$ and $\tau_C = 455(40)\mu s$) sides of the resonance, corresponding to the upper-left and lower-right branches of the dependence of the recoil frequency on the refractive index in Fig.2. In this case, the attenuation, proportional $\exp(-\tau^2/\tau_C^2)$, turns out to be noticeably less pronounced in the high-frequency part shown in Fig. 2 is the lower right branch of the dependence of the frequency (or energy) of the recoil depending on the magnitude of the deviation from the resonant frequency. This is probably due to the more distant location of the high-frequency branch from the dotted line in Fig.2, corresponding to the value of the refractive index $n=1$, compared with the low-frequency branch, which, as shown below, can only be explained using representations (15) and (17) of the Abraham theory.

Indeed, the indicated asymmetry of the location of the left and right branches of the curves relative to the dotted line cannot be explained within the framework of Minkowski theory, in which the same representation (8) for the photon pulse should be used for both the left branch (where $n>1$) and the right branch (where $n<1$).

According to (8), the curves of the dependence of the recoil frequency on the magnitude and sign of the deviation $\Delta/2\pi$ should be located at the same distance from the ordinate axis in Fig. 2, corresponding to the zero value $\Delta/2\pi = 0$ of the frequency deviation.

In this case, for a positive deviation value $\Delta/2\pi > 0$ in the case $n<1$ of a curve corresponding to Minkowski theory, it is represented as a continuous curve marked with an arrow with the letter M. This curve it no longer corresponds to experimental data in this area.

However, already on the basis of Abraham's theory, which leads to relation (17) in this frequency range, it is possible to interpret the asymmetry of the location of these two branches of the curves observed in [59] relative to the axis $\Delta/2\pi = 0$ in Fig.2 for $n>1$ and $n<1$.

Let's show this by considering, for simplicity, the data shown in Fig. 2 (see Fig.3 in [59]) at the same modulus values of the $\Delta/2\pi = \pm 250$ MHz frequency, which corresponds to the same modulus deviation from unity for the refractive index of the medium, according to the estimate $n-1 = \mp\delta; 0 < \delta \ll 1$ given in [59] (see formula (4) [59]).

In this case, the deviation of the recoil frequency for values $n = 1-\delta < 1$ at $\Delta/2\pi = 250$ MHz is about -1348 Hz, which is modulo about three times greater than the value of 552 Hz, following from Fig. 3 in [59] at $\Delta/2\pi = -250$ MHz, when $n = 1+\delta > 1$.

Indeed, the indicated frequency difference should be proportional to the magnitude of the difference $\Delta\omega \propto p_{rec}^2(n \neq 1) - p_{rec}^2(n=1)$ in the squares of the magnitude of the momentum received by the atom when it absorbs a photon. Taking into account (in the case $n = 1+\delta > 1$) the sum (15) of the Abraham-shaped photon momentum and the Abraham force impulse, we obtain that $\Delta\omega(n>1) \propto 2\delta + O(\delta^2); \delta \ll 1$. For the case $n = 1-\delta < 1$ of a similar value presented in (17), we obtain an estimate $\Delta\omega(n<1) \propto -6\delta$ and thus, indeed, the relation $|\Delta\omega(n<1)|/\Delta\omega(n>1) = 3$ is valid.

As noted above, within the framework of the Minkowski representation (8) or the use of relation (15), not only for the refractive index $n>1$, of the medium, but also for the values $n<1$, there is no longer a possibility of explaining the above asymmetry, since for the case $n = 1-\delta < 1$ of the frequency difference $\Delta\omega(n<1) = -2\delta$ and $|\Delta\omega(n<1)|/\Delta\omega(n>1) = 1$, respectively. This is illustrated by the curve in Fig. 2, indicated by the arrow with the letter M.

This is due to the fact that the Minkowski theory lacks the Abraham force, the momentum of which must be additionally taken into account in (15) along with the momentum of the photon itself in the medium not only at values $n>1$, but also at values $n<1$ according to (17).

Indeed, in [38] (see the footnote on page 321 in [38]), when the refractive index is $n<1$, the Minkowski representation for the photon momentum (11) formally coincides with the Abraham representation (13). However, Minkowski's theory lacks the Abraham force and the Abraham force momentum (16), which leads to the ratio (17) for the resulting recoil momentum, which no longer coincides with equality (10), which is valid only for values $n>1$.

### 4.3 Abraham's and Roentgen's forces

Analysis of the experimental data [59] was also carried out in [63] using not only Minkowski theory, but also the Abraham-shaped representation for the EMF pulse. The aim of [63] is to substantiate the possibility of "reconciling" the theories of Abraham and Minkowski, when both theories are considered "correct", but correspond to different forms of EMF description in the environment. However, as noted above, in principle, such simultaneous recognition of contradictory theories is impossible due to the experimentally established fact of the existence of the Abraham force in [6]-[8] and [45], [46]. Moreover, the Abraham force in [63] is not even mentioned.

It is noted in [63] that the key to solving the dilemma of the Minkowski and Abraham theories is to take into account the impulse density of the Roentgen force, which has the form [63] (see formula (19) in [63]):

$$\vec{g}^R = \frac{(n^2-1)}{4\pi c}[\vec{E}\times\vec{H}] \qquad (30)$$

However, this relation (30) given in [63] is exactly equal to the momentum density of the Abraham force, since according to (27) there is a strict equality:

$$\vec{g}^R = \int dt \vec{f}^A + const \qquad (31)$$

In [38] (see (13.21) in [38]), when obtaining the above representation (14) for the volume-integral momentum of the Abraham force, the volume integral of (30) or (31) is actually used (in the case of an integration constant equal to zero in (31)).

It was only noted in [63] that the value $\vec{g}^R/\rho$ is equal to the difference between the mechanical momentum $m\vec{u}$ of an individual atom and the canonical momentum $\hbar\vec{\nabla}S$, where $m$ - is the mass of the atom, $\vec{u}$ - is the flow velocity of the condensate atoms characterized by the phase $S$ and $\rho$ - is the number density of particles in the condensate (for example, $\rho = 0.2(2\pi/\lambda)^3 \approx 10^{14} cm^{-3}$ in experiment [59]). In [63], it is assumed that the sum of the $\rho\hbar\vec{\nabla}S$ with the EMF momentum density in Minkowski theory is equal (see (1) in [63], where it is noted that (1) corresponds to experimental data [59]) to the sum of the $\rho m\vec{u}$ with the EMF momentum density in the Abraham form (see equality (2) in [63]). The value of velocity $\vec{u}$ (see (3) in [63]) is precisely determined from the condition that such equality of total momentum densities is fulfilled.

If we explicitly write down the equality of the indicated total momentum densities containing the Minkowski and Abraham representations for the EMF pulse (that is, if we equate the relations (1) and (2) given in [63]), then the following equality is obtained:

$$\vec{g}^A + \vec{g}^R = \vec{g}^M \qquad (32)$$

Equation (32) uses the force impulse density (30) introduced in [63], as well as representations (24) and (29) for EMF momentum densities in the form of Abraham and Minkowski, respectively.

Note that equality (32) always holds identically for any values of the refractive index of the medium. In fact, it is precisely thanks to this identity (32) that the equality of representations (23) and (28) for the law of conservation of momentum in the form of Abraham and Minkowski, noted above and in [38], is realized.

The conclusion of [63] that representation (30) and the following identity (32) underlie the contradiction between the theories of Abraham and Minkowski is indeed true, although it is well known [38]. However, as noted above, after the discovery of the Abraham force, the equality of the Minkowski and Abraham theories considered in [63] contradicts the experimental results. [6]-[8], [45], [46], which reject Minkowski's theory [5].

Nevertheless, in [63], the momentum density of force (30) is considered only in connection with the mention of the well-known Roentgen force [73] acting from the side of a magnetic field on a dielectric moving in it. However, there is no in (30) any value of the velocity of the dielectric .

The representation for the Abraham force (27) was obtained in [38], on the contrary, precisely for the case of describing the effect of EMF on a stationary medium. However, in [38] the following generalization of representation (27) for the Abraham force density is given for the case of motion of a medium with velocity (see (13.39) in [38]):

$$\vec{f}^A = \frac{1}{4\pi c} \frac{\partial}{\partial t}\left[[\vec{D}\times\vec{B}] - [\vec{E}\times\vec{H}] + \frac{\vec{u}(\vec{u}([\vec{D}\times\vec{B}] - [\vec{E}\times\vec{H}]))}{c^2(1 - u^2/c^2)}\right] \qquad (33)$$

In (33) $\vec{D}$ and $\vec{B}$ - are the vectors of electric and magnetic induction, respectively. In this case, the EMF momentum in a moving medium already differs from the representation (24) and has the form in Abraham's theory [38]:

$$\vec{g}^A = \frac{1}{4\pi c}\left[[\vec{E}\times\vec{H}] + \frac{\vec{u}(\vec{u}([\vec{D}\times\vec{B}] - [\vec{E}\times\vec{H}]))}{c^2(1 - u^2/c^2)}\right] \qquad (34)$$

In the case of a nonmagnetic, isotropic, and stationary medium without dispersion, representations (33) and (34) transform into (27) and (24), respectively.

On the contrary, in the case when the condition of isotropy of the medium in Abraham's theory is violated, in addition to the density of the Abraham force (27), it

27is necessary to take into account the effect of an additional force (also absent in Minkowski's theory), which acts on a homogeneous, resting electrically neutral medium [1] (see footnote 9 in [1]):

$$\vec{f}_a^A = \frac{1}{8\pi} rot[\vec{D} \times \vec{E}] \qquad (35)$$

Moreover, only in the case of a medium at rest the relations $\vec{D} = \varepsilon\vec{E}$ and $\vec{B} = \mu\vec{H}$ are take place. However, in the case of slow non-relativistic velocities $\vec{u}$ of the medium, the following relations should be used instead [38] (see (13.27) in [38]) :

$$\vec{D} = \varepsilon\vec{E} + \frac{(\varepsilon\mu-1)}{\mu c}[\vec{u} \times \vec{H}] + O(u^2/c^2);$$
$$\vec{B} = \mu\vec{H} - \frac{(\varepsilon\mu-1)}{\mu c}[\vec{u} \times \vec{E}] + O(u^2/c^2) \qquad (36)$$

In the experiment [59], the condensate atoms are exposed to two short pulses of laser radiation with the duration of 5 microseconds, followed by each other with an interval of about 600 microns. After the first pulse, the atoms acquire a velocity of about 1.2 cm/sec. Therefore, when exposed to the second pulse of laser radiation, a situation does arise that can be characterized by the effect of the Roentgen force acting on a dielectric moving in a magnetic field, which is considered in [63].

Taking into account (36), the corresponding effect of Abraham's power (33) can also be estimated. Indeed, from representation (33), after substituting relations (36) into it, we obtain:

$$\vec{f}^A = \frac{(\varepsilon\mu-1)}{4\pi c} \frac{\partial}{\partial t}\left([\vec{E} \times \vec{H}] - \frac{1}{c}(\vec{u}(\mu H^2 + \varepsilon E^2) - \vec{H}\mu(\vec{u}\vec{H}) - \vec{E}\varepsilon(\vec{u}\vec{E})) + O(u^2/c^2)\right) \qquad (37)$$

At low atomic rebound rates in the experiment [59], as can be seen from (37), a change in the magnitude of the Abraham force density due to the motion of the medium can only be significant if the EMF energy is sufficiently high during repeated laser irradiation of the condensate.

Thus, the example given in section 3.2 of the analysis of experimental data [59] based on Abraham's theory shows the best agreement with these data precisely with Abraham's theory (with the obligatory consideration of Abraham's force and its momentum), and not Minkowski's theory, as was previously believed [60]-[63]. It is also established that the attempt to "reconcile" the theories of Abraham and Minkowski considered in [63] on the basis of the Roentgen force, in fact, boils down only to taking into account the momentum density of the Abraham force, when the formal coincidence in (15) of the conclusions of these theories takes place only under the condition $n > 1$.

On the contrary, for the case $n < 1$ corresponding to the experimental data [59] shown in Fig.2, as shown in Section 3.2, only the Abraham theory allows us to interpret these data based on the Abraham force, which is absent in the Minkowski theory.

**4.4 A contradiction in the analysis of the dilemma of the theories of Minkowski and Abraham**

In the well-known work of Brevik (1979) [43] and his subsequent recent works [60],[62], [67], [68], [70] the legitimacy of using the Minkowski representation for the EMF pulse in the medium is actively defended, despite the recognition in these works of the very existence of the Abraham force. In particular, it was noted in [43] that "the experiments of Walker et al. (1975) and James (1968) support the Abraham tensor in contrast to the Minkowski tensor." Although in such an implicit form, without mentioning the power of Abraham, it nevertheless means that these experiments discovered and measured the power of Abraham. At the same time, [43] does not even mention the main conclusion of these experiments, which unequivocally reject the Minkowski theory, in which there is no Abraham force for any EMF frequencies. At the same time, such an experimentally discovered example of the manifestation of the Abraham force equally unequivocally rejects the Minkowski representation for an EMF pulse in a medium, since the form of this representation can make physical sense only in the strict absence of the Abraham force in all possible frequency ranges of the EMF in the medium. Nevertheless, regarding the consideration of high-frequency EMFs in optical phenomena, [43] states: "In this case, direct observation of the fluctuating Abraham force is impossible. Under normal conditions, it is easiest to use the Minkowski tensor without divergence." In [60], this assumption is also presented in the form: "The Abraham force in the EMF momentum conservation equation simply disappears at high optical frequencies." However, in both [43] and [60], this statement about the disappearance of the Abraham force for high-frequency EMF is not supported by rigorous evidence of the complete absence of this force. Apparently, there can be no such proof, since even recognizing the very fact of the existence of a small and strongly fluctuating force does not mean that it can be recognized as identical to zero. Only in the case of a strictly zero Abraham force can it make sense to consider the EMF pulse in the Minkowski form.

Moreover, there is experimental evidence of the existence of a non-zero Abraham force in the field of optical frequencies [47]-[52].

Thus, there are quantitative experimental measurements of the Abraham force associated with a propagating optical wave. This force is released by a guided light wave passing through an adiabatic mode transformation along a liquid-filled hollow optical fiber. Using this distribution of light intensity in a liquid, a time-averaged Abraham force density is generated. The incident laser field causes a linear axial displacement of the air-liquid interface inside the HOF, which provides a direct experimental measurement of the Abraham force density. At the same time, a good agreement was obtained between the experimental results and the theoretical definitions of the Abraham force density.

In turn, these experiments are in addition to the previously obtained measurements of the Abraham force. [6]-[8], [45]-[46] They allow us to unambiguously reject Minkosky's theory and the corresponding representation for the EMF pulse in the medium used in the quantum VCR theory [23]-[25].

So, the simultaneous recognition of the physical reality of the Abraham force and the use of the Minkowski representation for the EMF momentum in the medium is internally contradictory and limits the application of Abraham's theory in analyzing observational data on the interaction of EMF and matter.
Therefore, it seems important to abandon the existing widespread application of Minkowski theory and instead use Abraham's theory to describe EMF in the medium.

## 5. Abraham's theory and the mass of a photon in a medium

The macroscopic theory of the Tamm-Frank VCR [22] and the quantum theory of the Ginzburg VCR [23] do not take into account the characteristic microscopic mechanism of the emission of the VCR quantum by the medium [1]-[4] and the threshold for the realization of the VCR effect (see Table 1).

As a result, based on the well-known classical and quantum VCR theory, it turns out to be problematic to describe the fundamental phenomenon of Cherenkov-type radiation, which is emitted precisely by a medium under the influence of so-called Bessel rays free from diffraction [74]-[88].

On the contrary, the quantum VCR theory [11]-[13], based on the theory of Abraham [9],[10], already takes into account the mechanism of emission of photons by the medium due to the fact that only in Abraham's theory it is possible to consider photon mass in the medium in the form (6).

In this section, the possibility of applying a new theory of VCR based on the theory of Abraham to describe the radiation initiated by Bessel rays is considered. In this regard, a general idea of these rays is given below, and then the grounds for introducing an effective photon mass are analyzed in more detail when considering the process of EMF propagation in the medium and in the waveguide (both in the presence and absence of a dielectric medium inside the waveguide).

The study of energy-efficient processes of propagation of EMF waves in a medium in the form of practically diffraction-free rays has well-known fundamental and applied significance [74]-[79]. These rays are sometimes also called Bessel rays [76] in contrast to ordinary Gaussian rays, for which diffraction determines a significant spatial scattering of the beam energy. At the same time, various mechanisms for creating such narrow cross-sectional beams are considered, the diameter of which is interconnected with the length of the electromagnetic wave, usually exceeding it by several orders of magnitude [74]-[79].

Of particular interest is the TGz radiation caused by such rays, which has numerous practical applications in medical diagnostics and other fields [78] (see also references in [78]). A similar radiation with characteristic properties resembling Cherenkov radiation is also observed during propagation in waveguides, which also provide spatial localization and concentration of electromagnetic wave energy [80]-[88].

To date, however, the mechanism of such radiation caused in the medium by the propagation of localized rays of electromagnetic waves is not fully understood, based on the generally accepted VCR theory [22]-[25], [38].

Indeed, the observed radiation in the form of a Cherenkov cone is caused not by a fast charged particle, but by a Bessel beam or their analogues. These rays propagate in the form of a plasma filament localized in the transverse direction [78], which plays the role of a waveguide, and their phase velocity may not exceed the phase velocity of the radiation they generate, which is unacceptable in the framework of the theory of microwave [22]-[25], [38].

As a result, various modifications of the theories are proposed. [76], [78], [79], [80]-[82], which are forced to use a number of additional assumptions for the possibility of explaining these effects on the basis of the well-known VCR theory [22]-[25], [38].

It is important that, according to observations, it is the medium itself, the polarization of which is caused by a localized beam that is the direct source of Cherenkov-type radiation. [77], [78], [80]-[82]. This is also a characteristic feature of the microscopic mechanism of IVF, as noted in [1]-[4] and [11]-[13].

This section shows how the radial Bessel localization of a beam can be related to a certain effective mass of a photon. At the same time, the objections raised in [44] in connection with the concept of the mass of a photon in a medium introduced in the theory of the microwave radiation [11]-[13] are considered and eliminated

## 5. Photon mass in the medium and in the waveguide

The idea of the finite nonzero mass of a photon, which a photon can have when propagating in any medium with a refractive index other than unity, was used in [11]-[13], [39], [40] and it requires a separate explanation, taking into account the existing observations about the mass in general and about the mass of the photon in particular [89]-[91].

Indeed, it probably makes sense for a photon to talk about its non-zero mass only in the context of its interaction with a medium in which an electromagnetic wave is already propagating at a speed different from the speed of light in a vacuum. Only in this case is it possible to introduce the idea of a nonzero real mass of a photon. It is usually assumed that the frequency of an electromagnetic wave and the corresponding photon energy in a refractive medium coincide with their values in a vacuum, in contrast to the wavelength and momentum of the photon [89], [90].

As a result, when the momentum changes, the photon must inevitably acquire a finite nonzero mass during its transition from a vacuum to a medium with a refractive index other than unity.

This should be the case due to the well-known relativistic relation (5) between the energy of a particle, its momentum and mass given in the Introduction [89], [90].

Relation (5) leaves the mass invariant during the transition from one inertial coordinate system to another, when the energy and momentum of the particle are transformed according to the Lorentz transformations as some four vector. In this case, the transition from one inertial system to another should be distinguished from the transition of a photon from a vacuum to a medium with a refractive index other than unity. Because in the presence of such an environment, there is always a dedicated coordinate system in which the environment, as a whole, is at rest.

Let us consider the question of the mass of a photon in a medium in connection with the opposite notion of zero photon mass even in a medium with a refractive index different from unity [44].

Here we present the argument [44], which is based on the use of the Klein-Gordon-Fock equation for a free vector relativistic particle with mass $m$ (see, for example, [44] and the references given there):

$$\left[\Delta - \frac{1}{c^2}\frac{\partial^2}{\partial t^2} - \frac{m^2 c^2}{\hbar^2}\right]\Psi_k = 0; k = 0,1,2,3 \qquad (38)$$

In equation (38) $\Delta$ - is the three-dimensional Laplace operator, and $\hbar$ - is Planck's constant. Equation (38) at the quantum level provides a relationship between the energy and momentum operator of a particle, regardless of its internal degrees of freedom, and is valid, for example, for describing bosons [44]. Equation (38), taking into account the representation for the energy and momentum operators in the form e $E = i\hbar\frac{\partial}{\partial t}; \vec{p} = i\hbar\vec{\nabla}$, is a direct consequence of the relativistic relation (5), written in the form виде $\frac{E^2}{c^2} - p^2 - m^2 c^2 = 0$ [89], [90].

According to the argument given in [44], for photons in a medium, the analog of the wave function in equation (38) is a vector potential satisfying the usual wave equation, which is compared with (38) and has the form:

$$\Delta\vec{A} - \frac{n_0^2}{c^2}\frac{\partial^2 \vec{A}}{\partial t^2} = 0 \qquad (39)$$

Thus, in [44] concludes that since (39) lacks an analogue of the mass term present in equation (38), the mass of a photon in a medium is zero in any medium having a refractive index $n_0 \neq 1$.

However, such a conclusion is made based on the implicitly assumed spherical symmetry of the solution of equation (39), when there is no distinguished direction corresponding, for example, to the direction of motion of a photon emitted by a polarized medium during the implementation of the VCR effect. In the presence of such a violation of spherical symmetry and its change to cylindrical symmetry with the axis of the cylinder coinciding with the direction of propagation of the Cherenkov photon, equation (39) should already be considered in the appropriate modification, which already resembles equation (38) with a nonzero mass of the photon.

For example, it may be shown that non-zero photon mass arising in all cases where an electromagnetic wave propagates in a waveguide or has a different reason for the localization of the wave field, such as for Bessel rays [74]-[76].

For simplicity, as in [74], instead of (39), we consider the wave equation for the scalar potential of an electromagnetic field in a medium with an index of refraction $n$, which has the form:

$$\Delta \Phi - \frac{n_0^2}{c^2} \frac{\partial^2 \Phi}{\partial t^2} = 0 \qquad (40)$$

The value of the refractive index $n_0$ of the medium introduced in (39) and (40) may, in the case of EMF propagation through a waveguide filled with a medium with a refractive index $n_0$, differ from the refractive index $n$ of an unbounded medium introduced in the previous sections.

We will search for a solution to (40) in an unbounded space by considering a cylindrical coordinates $(z, r, \varphi)$, where the axis $z$-is directed along the axis of a localized Bessel ray, for the field of which we use the well-known axisymmetric representation [74]:

$$\Phi(z;r;t) = \overline{\Phi}(z;t) J_0(k_\perp r) \qquad (41)$$

In (41) $J_0$-is a cylindrical Bessel function of order zero. After substituting (41) into (40), we obtain an equation of the type of the modified one-dimensional Klein-Gordon-Fock equation (38):

$$\frac{\partial^2 \overline{\Phi}}{\partial z_1^2} - \frac{1}{c^2} \frac{\partial^2 \overline{\Phi}}{\partial t^2} - \frac{k_\perp^2}{n_0^2} \overline{\Phi} = 0; z_1 = n_0 z \qquad (42)$$

From a comparison of equations (42) and (38), we obtain that for Bessel rays, the mass of a photon admits the following representation for any values of the refractive index $0 < n_0 \leq 1; n_0 \geq 1$:

$$m = m_{ph} = \frac{\hbar k_\perp}{n_0 c} \qquad (43)$$

A similar possibility for introducing an effective finite mass of a photon during the propagation of an electromagnetic wave in a waveguide was noted by B. M. Bolotovsky (private communication), when discussing the materials of the work [11] before its publication. The idea of the photon mass associated specifically with the propagation of EMF through a waveguide is indeed well known and, for example, was noted in Feynman lectures on physics [91] (see the analogy (24.31) with (24.17) in the vol. 2 of [91], which gives the relation (43) for the case $n_0 = 1$).

According to (43), the magnitude of the photon mass will be determined only by the waveguide size $l_\perp = \pi/k_\perp$ (see (24.13) in [91]) or by the characteristic size of the transverse localization of the beam, both in vacuum $n_0 = 1$ and in any medium $n_0 \neq 1$.

However, in [91] it is noted only the curiosity of this analogy between the dispersion equation describing the propagation of EMF through a waveguide and the relation (5) relating energy, momentum, and mass of a particle, leading to equation (38).

Moreover, explicitly the relation (43), as well as its special case with a value $n_0 = 1$, does not given in Feynman's lectures [91], which were published before 1975, and therefore there is no mention of Abraham's theory and the EMF pulse in this theory.

The presence in Abraham's theory of the actual mass of a photon, defined in the form (6) for any unlimited medium with a refractive index преломления $n \neq 1$ other than unity, additionally allows us to use the analogy noted in [91], leading to an estimate of the mass of a photon in a waveguide in the form (43).

For example, from the condition of equality of the photon mass defined in (43) with the photon mass (6) in Abraham's theory, it follows that the propagation of EMF through a waveguide is equivalent to the propagation of EMF in an unlimited medium with some refractive index other than unity, the value of which, in the general case, may differ from the value $n_0$ in (43).

Conversely, the simulation of photon propagation in any unlimited medium with a refractive index $n \neq 1$ can be based on the description of EMF propagation through a waveguide, the transverse size of which is determined from the condition of equality of the photon mass in (43) and in (6) and has the form:

$$l_\perp \equiv \frac{\pi}{k_\perp} = \frac{\hbar \pi c}{E_{ph}} F(n_0;n)$$

$$F = \frac{n}{n_0\sqrt{n^2-1}}, n>1; F = \frac{1}{n_0\sqrt{1-n^2}}, n<1 \quad (44)$$

In particular, for the case of a waveguide not filled with a medium, considered in [91], the condition $n_0 = 1$ must be used in (44).

It follows from (44) that the size of the transverse localization can be on the order of the wavelength of the EMF photon only at values of the refractive index of the medium markedly different from unity, and at values close to unity it can significantly exceed the wavelength of the EMF.

Moreover, in the limit $n \to 1$, the propagation of a photon in the medium ceases to have a definite direction, since the transverse size of the waveguide simulating this process increases indefinitely according to (44) $l_\perp \to O(\frac{1}{\sqrt{|n-1|}}) \to \infty$.

In this limit, the mass of a photon in medium (6) also has a zero limit. Thus, if we assume that the photon mass in Abraham's theory (6) is equal to the photon mass defined in (43) and corresponding to the directional propagation of the EMF along the waveguide, then we can talk about the acquisition of a finite mass by a photon in the medium only if the spherical symmetry characterizing the direction of the EMF beam in the medium is violated. Such a symmetry breaking occurs in the case of photon emission by the medium in the VCR effect.

However, in the general case, the value of the photon mass determined in (43) may not strictly coincide with the mass of the photon determined in (6) based on Abraham's theory and relations (4) and (5). Nevertheless, as shown below, even in the general case, the value (43) is superimposed The limitation from below is in the form of the mass value (6), which follows from the natural limitation on the magnitude of the group velocity of EMF waves (it should not exceed the speed of light in a vacuum) propagating through the waveguide.

Indeed, the presence of a finite mass (43) associated with the transverse size of the beam affects the magnitude of the phase and group velocity traveling along the axis of the Bessel beam, described by the solution of equation (42).

This is also typical for the observed [74], [78], [79] localized rays in an unlimited space and for traveling wave propagation in thin waveguide films [75], [77], [81]-[83]. Indeed, let the traveling wave amplitude in (42) have a representation $\overline{\Phi}(z,t) = \Phi_0 \exp(i(\omega t - k_z z))$ in which, according to (43), we have a dispersion equation in the form:

$$\omega = \frac{c}{n_0}\sqrt{k_z^2 + k_\perp^2}, \qquad (45)$$

$$k_z = \frac{2\pi n_0}{\lambda_0}\sqrt{1 - \frac{m_{ph}^2 c^2 \lambda_0^2}{4\pi^2 \hbar^2}}$$

In (45) the relations $k^2 = k_z^2 + k_\perp^2; (k_0 = \omega/c; k = k_0 n_0; \lambda_0 = 2\pi/k_0)$ are used.

According to this dispersion equation, the longitudinal, along the propagation of the beam, wavenumber $k_z$ is related in (45) to the length $\lambda_0$ of the electromagnetic wave in a vacuum (using the definitions $k^2 = k_z^2 + k_\perp^2; k_0 = \omega/c; k = k_0 n_0; \lambda_0 = 2\pi/k_0$).

The expression for the phase velocity of the wave according to (45) has the form:

$$V_{ph} = \frac{\omega}{k_z} = \frac{c}{n_0}\left(1 - \left(\frac{\lambda_0}{\lambda_{ph}}\right)^2\right)^{-1/2}; \lambda_{ph} = \frac{2\pi\hbar}{m_{ph}c} \qquad (46)$$

The group velocity, according to (45), for the case of the absence of dispersion with a medium refractive index $n_0$ that is independent from the wavenumber has the form:

$$V_g = \frac{\partial \omega}{\partial k_z} = \frac{c}{n_0}\left(1 - \left(\frac{\lambda_0}{\lambda_{ph}}\right)^2\right)^{1/2} \qquad (47)$$

The next relationship follows from (46) and (47):
$$V_{ph}V_g = c^2/n_0^2 \tag{48}$$

As follows from (47), for a medium without dispersion with a refractive index value $n_0 > 1$, the value of the group velocity for any photon mass in (43) or the corresponding transverse beam size is always less than the speed of light in vacuum $V_g < c$.

However, for refractive index values $n_0 < 1$, the inequality $V_g < c$ for the value of the group velocity from (47) is fulfilled only with the following restriction on the mass of the photon and on the transverse size of the beam related to it by the ratio (43) in the case when the refractive index in the waveguide is the same as in the free space, like in (6) $n_0 = n$ (this case is also considered below):

$$\begin{aligned} m_{ph} &\geq m_{ph}^A(n<1) = \frac{E_{ph}}{c^2}\sqrt{1-n^2}; \\ l_\perp &= \frac{\pi}{k_\perp} \leq \frac{\pi \hbar c}{E_{ph}n\sqrt{1-n^2}}; \\ E_{ph} &= \hbar\omega = \hbar k_0 c = \frac{2\pi\hbar c}{\lambda_0} \end{aligned} \tag{49}$$

At the same time, in (49), as usual [90], it is assumed that the energy of a photon in a medium coincides with its energy in a vacuum.

If the phenomenon of refractive index dispersion is taken into account, the type of phase velocity obtained in (46) and the known relationship $V_{ph}V_g = c^2$ between phase velocity and group velocity should be used to determine the group velocity [92]. This leads to the following form for the group propagation velocity along the axis $z$ of a localized Bessel beam:

$$V_g = cn\sqrt{1-\left(\frac{\lambda_0}{\lambda_{ph}}\right)^2} \tag{50}$$

Since the group velocity cannot exceed the speed of light in a vacuum, for the case of plane wave propagation in the limit $\lambda_0/\lambda_{ph} = k_\perp/k_z \to 0$, expression (50) always makes sense only for the case $n < 1$ considered in [92].

For the case $n > 1$ from the condition $V_g \leq c$ from (50), we obtain the following constraints for the mass of the photon and the corresponding transverse size of the localization of the Bessel beam (41) in a medium with the refractive index dispersion:

$$m_{ph} \geq m_{ph}^A(n>1) = \frac{E_{ph}\sqrt{n^2-1}}{c^2 n}; n>1$$

$$l_\perp = \frac{\pi}{k_\perp} \leq \frac{\pi \hbar c}{E_{ph}\sqrt{n^2-1}} = \frac{\lambda_0}{2\sqrt{n^2-1}} = l_{\perp\max} \qquad (51)$$

$$E_{ph} = \hbar\omega = \hbar k_0 c = \frac{2\pi\hbar c}{\lambda_0}$$

Thus, in conditions (49) and (51), the threshold value limiting the mass of a photon from below exactly coincides with the definition of the mass of a photon in a medium with an arbitrary refractive index, which is obtained from definition (5) for the mass of any relativistic particle when using the Abraham representation (4) for the momentum of a photon in a medium [38].

Indeed, if we use relation (5), which underlies equation (38), in the case of describing the photon momentum in a medium in the form of Abraham (4), then for any values of the refractive index other than unity (equal to unity only in vacuum), we can determine the final actual mass of the photon in the medium in the form of (6).

The value of the mass of a photon in a medium, defined in (6), is a relativistic invariant quantity, which is a direct consequence of the relativistic invariance of relation (5). In (6), the mass of a photon remains a real quantity for the case $n<1$. Apparently, attention was not paid to this in [44], where the opposite statement is made about the imaginary of the photon mass at values in the case $n<1$ of using representation (4) in (5) for the photon momentum in an Abraham-shaped medium

## 5.2 The Bessel rays

Let us consider the examples of application of the condition (51) to the Bessel rays and in the non-linear optics.

The first example corresponding to the results of observing a Bessel beam at $\lambda_0 \approx 0.632 \times 10^{-6}$ m, when the diameter of the observed beam cross-section had an estimate of $l_\perp \approx 24.5 \times 10^{-6}$ m, following from the formula $l_\perp = \lambda_0(f/\pi d + O(d/f)); d \ll f$ given in [74] for the focal length of the lens $f = 0.305$ m and the diameter of the interference slit $d = 0.0025$ m. For the maximum distance over which a Bessel beam can propagate without noticeable changes in its peak intensity, an estimate of $Z_{\max} = 2\pi R l_\perp / \lambda_0 \approx 0.85$ m is given in [74] when the lens aperture radius is $R = 0.0035$ m. For comparison, in the case of a conventional Gaussian beam, the peak intensity decreases by an order of magnitude due to diffraction effects after passing through the characteristic Rayleigh length $Z_R = \pi d_\perp^2 / \lambda_0 \approx 0.05$ m.

Note that if, in an experiment similar to [74], the effects of refractive index dispersion of the medium could be significant, then condition (51) for the case of propagation of a Bessel beam in air yields the following estimate of the maximum propagation length without diffraction loss $Z_{max} \leq 2\pi R l_\perp^A / \lambda_0 = R/\sqrt{n^2-1} \approx 0.1506 \text{m}$. In this case, the condition $Z_{max} \gg Z_R$ still turns out to be fulfilled, since for the characteristic Rayleigh length we have an upper bound $Z_R \leq \pi (l_\perp^A)^2 / \lambda_0 = \lambda_0 / 4\pi(n^2-1) \approx 93.2 \times 10^{-6} \text{m}$.

For the large amplitude of the electric field $|\vec{E}_0| > E_{0th}$ in a beam of light propagating in a transparent medium, due to nonlinear effects associated with an increase in the refractive index $n = n_0 + \Delta n; \Delta n = n_2 |\vec{E}_0|^2$ in the beam propagation zone, it self-focuses, similar to the action of a refractive lens in the previous example when forming a Bessel beam. [77], [78], [83], [84]. In this case, the effects of the dispersion of the refractive index of the medium during beam filamentation [77], [78] are already significant and similar estimates based on (51) make sense.

Let us consider the propagation of a beam in a cylindrical waveguide formed during self-focusing, the axis of which is directed along the cylindrical coordinate $z$. In this case, in order to implement self-focusing, it is necessary to fulfill the following restrictions on the transverse wavenumber $k_\perp$ and the corresponding transverse beam size $l_\perp = \pi/k_\perp$, which determine the angle of inclination of the beam direction relative to the axis of the waveguide [80]:

$$\theta = \arccos\left(\frac{k_\perp}{\sqrt{k_z^2 + k_\perp^2}}\right) \leq \theta_0 = \arccos\left(\frac{n_0}{n_0 + \Delta n}\right) \tag{52}$$

Accordingly, only rays with an inclination to the axis $z$ satisfying condition (52) can return to the axis due to the difference in the refractive index $n_0$ of the medium outside the localization of the beam from the refractive index $n_0 + \Delta n$ in the localization region only in the case of $\Delta n > 0$.

It follows from (52) that there is the following restriction from below on the radius of the cylindrical waveguide channel:

$$l_\perp = \frac{1}{k_\perp} \geq \frac{1}{k_z}\left[\left(\frac{n_0 + \Delta n}{n_0}\right)^2 - 1\right]^{-1/2} = l_{\perp \min} \tag{53}$$

Condition (51), by giving an upper limit on the channel size, together with condition (53) makes it possible to determine the appropriate conditions for the magnitude of the refractive index fluctuation to realize localized beam propagation from the ratio $l_{\perp \max} > l_\perp > l_{\perp \min}$. In the limit $\lambda_0 / \lambda_{ph} = \lambda_0 / \pi n l_\perp \ll 1$, from the condition of compatibility of the left and right inequalities in this ratio, we obtain an estimate:

$$\left(\frac{n_0 + \Delta n}{n_0}\right)^2 > 1 + \frac{(n_0^2 - 1)}{n_0^2}\left(1 + \frac{\lambda_0^2}{\lambda_{ph}^2} + O(\frac{\lambda_0^4}{\lambda_{ph}^4})\right) \tag{54}$$

In particular, for the above example of Bessel rays observed in [74], the value is $\lambda_0 / \lambda_{ph} \approx 0.008/n_0$ in (54). This simplifies the dependence on the value $n_0$ for the lower limit obtained from (54) on the magnitude $\Delta n$ of fluctuations in the refractive index, at which such localized rays can be realized.

For simplicity, when obtaining estimate (54), it is assumed that an approximate substitution $n = n_0 + \Delta n \to n_0$ is used in the expression for the threshold beam size $l_{\perp \max}$ that is given in (51).

For a beam having a flat front when entering a nonlinear medium, the angle $\theta = \theta_d = \dfrac{0.61\lambda_0}{2l_\perp n_0}$ is determined by the diffraction effect [80]. In this case, nonlinear refraction completely compensates for the diffraction effect and the beam propagates without disturbing its radial localization, when the condition $\theta = \theta_d = \theta_0$ is provided. From this condition, we obtain an estimate for the transverse size of the waveguide cylinder:

$$l_\perp = \frac{0.61\lambda_0}{2n_0 \arccos\left(\dfrac{n_0}{n_0 + \Delta n}\right)} \tag{55}$$

It follows from (55) and constraint (51) that in addition to constraint (54) from below for the magnitude of fluctuations in the refractive index, there is an additional constraint from above, which has the form:

$$\cos^{-2}\left(\frac{0.61\pi}{n_0}\sqrt{n_0^2 - 1}\right) > \left(\frac{n_0 + \Delta n}{n_0}\right)^2 \tag{56}$$

Taking into account the fact that the Compton wavelength of a photon in a medium also depends not only on the photon energy, but also on the refractive index, the condition of compatibility of inequalities (54) and (56) imposes an additional restriction on the value of the refractive index of the medium. It depends on the energy of the photon and must be performed to realize the self-focusing of light in the medium. In particular, for the example of Bessel rays observed in [74] in air with $n_0 = 1.00027$, this condition for the compatibility of inequalities (54) and (56) is fulfilled, since the left part of (56) is approximately 1.00198, and the right part of inequality (54) is approximately 1.00054.

Thus, when considering localized Bessel rays, natural restrictions (49) or (51) arise on their transverse ray size from above for media without dispersion at values $n < 1$ and for media with refractive index dispersion at values $n > 1$ respectively. At the same time, it turned out that the threshold values for the corresponding transverse wave numbers and the values of the photon mass determined on their basis in (43) exactly coincide with the mass of a photon determined in (6) in a medium having an impulse (4) of the Abraham theory [38].

## 6. A new quantum microscopic theory of VCR emission by a medium

In the Introduction, it is noted that the new quantum VCR theory [11]-[13], based on the theory of Abraham and therefore taking into account the well-known characteristic microscopic mechanism of the emission of a quantum of VCR by the medium, gives a more accurate estimate of the threshold of VCR realization in (2) than the well-known VCR theory of Tamm-Frank [22] and Ginzburg [23] in (1) (see Table 1)

Therefore, it seems necessary to take into account this difference between the estimates of the threshold and representation (2) in many applications based on the use of estimate (1), for example, in connection with the problem of identification of relativistic particles in Cherenkov detectors [17],[18] and in the design of the accelerator equipment protection system [19], [20].

However, before considering the modification of these applications of the VCR effect based on the new quantum theory of VCR, it seems necessary to give the conclusion of condition (2) for comparison with the similar conclusion (1) from relation (18) obtained in the quantum VCR theory of the Ginzburg [23], [38], [4].

## 6.1 The momentum and energy conservation when VCR is emitting by medium

The new quantum theory of photons [11]-[13] uses the Abraham representation for the photon pulse in a medium in the form (4) (see also (12), (13)).

The energy and momentum balance equations used below take into account the change in the energy $\Delta E = E_0 - E_F = \Delta M c^2 = m_{ph}^A c^2 > 0$ of the medium (7), which is determined by the value (6) of the rest mass $m_{ph}^A$ of a photon in the medium and is caused by the emission of a quantum by the medium [11]:

$$m_e c^2 \gamma_0 + Mc^2 = (M - \Delta M)c^2 \gamma_2 + E_{ph} + m_e c^2 \gamma_1;$$

$$m_e \vec{V}_0 \gamma_0 = (M - \Delta M)\vec{V}_2 \gamma_2 + \frac{E_{ph}\vec{V}_{ph}}{c^2} + m_e \vec{V}_1 \gamma_1; \qquad (57)$$

$$\gamma_\alpha = \left(1 - \frac{V_\alpha^2}{c^2}\right)^{-1/2}, \alpha = 0;1;2$$

$$\vec{P}_{ph} = \frac{\vec{V}_{ph}}{c^2} E_{ph}; V_{ph} = \frac{c}{n}; n > 1; V_{ph} = nc; n < 1$$

In (57) $M$ - is the mass of the medium before the medium emits the VCR photon in any isotropic medium with a refractive index $n \neq 1$.

Equations (57) describe the balance of energy and momentum, which correspond to the condition of emission of a Cherenkov photon directly by the medium itself. In this case, the polarization of the medium necessary for the creation of a coherent VCR by the medium is the result of the interaction of the medium with a sufficiently fast-moving charged particle according to the well-known concept of the microscopic mechanism of the VCR [1]-[4].

In fact, equations (57) describe a change in the state of the entire closed system consisting of a medium, a moving charged particle and a Cherenkov photon immediately at the moment of the birth of this photon by the medium.

Note that the quantum VCR theory of the Ginzburg [23], [38] uses similar laws of conservation of energy and momentum, but using instead of (4) the Minkowski representation (8) for the momentum of a photon in a medium.

As noted in section 3, in the quantum VCR theory of the Ginzburg [23], as well as in the macroscopic VCR theory of the Tamm-Frank [22], the analog of the balance equations (57) refer to a time sufficiently remote from the moment when the Cherenkov photon is emitted by the medium, when a coherent VCR field has already been established in the medium.

It is clear from Fig.1 that the interference maximum characteristic of coherent radiation in the cone does not coincide with the cone boundary.

According to Table 1, the observed boundary of the VCR cone and the threshold of the VCR effect cannot be determined within the framework of the Tamm-Frank and Ginzburg theories of the VCR from relation (1).
The threshold velocity of a charged particle corresponding to the boundary of the VCR cone, however, can already be established in the form (2) precisely due to the consideration of the above-mentioned microscopic mechanism of Cherenkov photon generation directly based on the energy and momentum balance equations in the form (57).

Thus, from the balance equations (57), which take into account the change in the energy of the medium $\Delta E = E_0 - E_F = m_{ph}c^2 > 0$, the following conditions for the emission of coherent microwave radiation by the medium are obtained

$$1 \geq \cos\theta = \frac{c}{V_0 n_*}\left[1 + \frac{\varepsilon}{\gamma_0}\begin{cases}\frac{\sqrt{n^2-1}}{n}, n>1 \\ \sqrt{1-n^2}, n<1\end{cases}\right]; \gamma_0 = \frac{1}{\sqrt{1-\frac{V_0^2}{c^2}}}; \varepsilon = \frac{\varepsilon_{ph}}{m_e c^2}, \quad (58)$$

$$n_* = \begin{cases}n + \sqrt{n^2-1}, n>1 \\ \dfrac{1+\sqrt{1-n^2}}{n}, n<1\end{cases} \quad (59)$$

$$V_0 > V_{th} = \frac{c}{n_*}\left[1 + \frac{\varepsilon}{\gamma_0}\begin{cases}\frac{\sqrt{n^2-1}}{n}, n>1 \\ \sqrt{1-n^2}, n<1\end{cases}\right] \quad (60)$$

Conditions (58) and (60) give condition (2) sufficient for the occurrence of the VCR and, thus, it is possible to obtain the maximum angle $\theta_{max}$ of the VCR cone of the VCR rays, which is consistent with what was observed in the experiment (see Table 1 in the Introduction).

In the case of an isotropic plasma with a refractive index $n<1$, the VCR theory of the Frank-Tamm and Ginzburg, as is known [38], completely excludes the possibility of realizing the VCR effect in the form of transverse high-frequency electromagnetic waves.

The new VCR theory, based on the theory of Abraham, provides such a possibility, which is described by the threshold condition (2) following from (58).

Note that representation (58) is qualitatively different from condition (18) obtained in Ginzburg's quantum theory [23], [38].

In the final two sub-sections of this section, a more detailed quantitative comparison of the conditions for the implementation of the VCR (2) and (1) is carried out, considering the case when the refractive index of an isotropic medium is very close to unity. It would seem that in this case there should be a small difference in the conclusions based on the Minkowski and Abraham concepts for the photon momentum in medium.

However, this is not the case, and the differences in estimates based on (1) and (2) for the threshold value of the parameter $\gamma = (1 - V^2/c^2)^{-1/2}$ can be several orders of magnitude precisely in the limit of small quantities $n-1 \ll 1$ precisely because of the consideration in (2) of the change in the energy of the medium in the form of (7) when the medium emits a quantum of the VCR.

## 6.2 A relativistic generalization of the Landau criterion and the generation of a photon by a medium

Let us consider an analogy between the microscopic mechanism of threshold emission of a quantum by a medium described by condition (2) and the Landau criterion (1941) [32] for the destruction of a superfluid state. In [11], a relativistic generalization of this criterion was obtained in the form of the IVH condition (2), in which the inequality sign should be considered, in contrast to the equality sign following from condition (58) established on the basis of the quantum theory of IVH using the laws of conservation of energy and momentum.

Indeed, in the theory of Landau (1941) [32], the threshold mechanism of the generation of elementary vortex excitation is associated with the dissipative instability of the system, due to its lack of conservativeness due to the presence of extremely low, but not strictly zero viscosity. In [32], the motion of liquid helium through a thin capillary is considered, which, due to the low-temperature coherent superfluid state of helium, is carried out at a constant velocity $U_s$ along the axis of the capillary directed along the axis $z$.

At the same time, it was noted in [32] that in this state, during the movement of helium, the effect of viscosity does not actually manifest itself, which, due to friction against the walls of the capillary and friction inside the liquid itself, would dissipate the kinetic energy of the liquid's motion and its gradual deceleration.

However, it turns out that such a coherent superfluid state can become unstable and be destroyed if the constant velocity $U_S$ begins to exceed a certain threshold value, as follows from experimental data and according to the Landau theory [32].

To obtain an estimate of the magnitude of this threshold velocity, [32] introduces the idea of elementary disturbances (phonons, rotons, vortex rings), threshold, energetically favorable, the generation of which near the walls of the capillary is possible only if any arbitrarily small but not equal to zero viscosity is taken into account.

Since it is important to establish the magnitude of the energy $E_0$ and momentum $\vec{P}_0$ of these perturbations, determined precisely in a stationary liquid, in [32] in this regard it is proposed to consider the flow of liquid helium in a frame of reference moving with helium, in which the helium is at rest, and the walls of the capillary move with velocity $-U_S$.

In this regard, it is noted in [32]: " In the presence of viscosity, the resting helium should also start moving. It is physically obvious that due to the interaction with the moving walls of the tube, the movement of the entire liquid as a whole cannot occur from the very beginning. The appearance of motion should begin with the excitation of internal movements in the liquid layers close to the wall, i.e. with the excitation of phonons or rotons in the liquid."

After these elementary excitations appeared in one form or another in the liquid, it was proposed in [32] to return to the laboratory frame of reference with a fixed capillary wall. This is necessary to compare the initial value $E_I = M_I U_S^2 / 2$ of the kinetic energy of the flow, which characterizes the flow before the birth of disturbances in it, with the energy of the flow $E_F = E_1 + M_F U_S^2 / 2$ after the birth of a disturbance with energy $E_1 = E_0 - P_{0z} U_S$ [32]. Here, the non-relativistic limit $U_S \ll c$ of the following well-known energy conversion formula was used in [32] for the perturbation energy $E_1$ in the laboratory frame of reference [91] (see (17.10) in vol. 1 in [91]):

$$E_1 = \frac{E_0 - P_{0z} U_S}{\sqrt{1 - U_S^2 / c^2}} \qquad (61)$$

In [32], it is implicitly assumed that the mass of a liquid entrained by an elementary disturbance is negligible when it is acceptable to consider the limit $\Delta M = M_I - M_F \to 0$ and the corresponding equalities $M_I = M_F = M$.

As a result, the following Landau criterion for instability and destruction of a coherent superfluid is obtained in [32] from the requirement of fulfilling the inequality $E_I > E_F$, which means that after the generation of disturbances, the energy of the flow cannot increase, but should only decrease at nonzero viscosity:

$$E_1 = E_0 - P_{0z}U_S < 0 \qquad (62)$$

In (62), the magnitude of the projection of the elementary excitation pulse on the direction of the flow velocity along the capillary axis can be expressed as $P_{0z} = |\vec{P}_0|\cos\theta; P_0 \equiv |\vec{P}_0|$, where $\theta$ - is the angle between the direction of propagation of the elementary disturbance and the capillary wall. Therefore, (62) implies the representation of the Landau criterion in the form:

$$1 \geq \cos\theta > \frac{U_{th}}{U_S};$$

$$U_{th} = \frac{E_0}{P_0} \qquad (63)$$

$$U_S > U_{th} \qquad (64)$$

In particular, the Landau criterion was considered in [32] using the example when an elementary disturbance generated due to nonzero viscosity near the capillary wall is a phonon.

The phonon corresponds to the well-known dispersion law $E_0 = c_S P_0$, for which it follows from the Landau criterion (64) that the superfluid state of liquid helium flowing almost frictionlessly through a capillary becomes unstable and collapses at a helium flow velocity exceeding the speed of sound $U_S > U_{th} = c_S$.

In [11], a condition for the implementation of the VCR effect was obtained in the form of inequality (2), which is the result of a relativistic generalization of the Landau criterion, obtained in [32] in the form of (62)-(64) to describe the dissipative instability of the flow of superfluid helium.

We present the conclusion of inequality (2) obtained in [11]. Note that the energy balance equation (57) describing the generation of a photon by medium is written in the reference of frame in which the medium is at rest, and the electron moves at a constant initial velocity $\vec{V}_0$ before the photon is emitted by medium, and after the photon is emitted, its velocity differs from the initial velocity and is equal to $\vec{V}_1$.

By analogy with the Landau theory [32], we will consider the electron as a prototype of the capillary walls. Therefore, let's imagine the energy balance equation (57) in the reference of frame moving at a constant speed $\vec{V}_0$.

In this frame of reference, the electron is at rest, and the medium moves with velocity $-\vec{V}_0$, and the energy balance equation in (57) has the form [11] (see (5) in [11]):

$$m_e c^2 A(\vec{V}_0; \vec{V}_1) = \varepsilon_{ph} \gamma_0 B(\vec{V}_0; \vec{V}_{ph});$$

$$A \equiv 1 - \gamma_0 \gamma_1 \left(1 - \frac{(\vec{V}_0 \vec{V}_1)}{c^2}\right); \gamma_0 = \left(1 - \frac{V_0^2}{c^2}\right)^{-1/2}; \gamma_1 = \left(1 - \frac{V_1^2}{c^2}\right)^{-1/2}; \quad (65)$$

$$B \equiv 1 - \frac{(\vec{V}_0 \vec{V}_{ph})}{c^2} - \sqrt{1 - \frac{V_{ph}^2}{c^2}};$$

$$\vec{P}_{ph} = \frac{\vec{V}_{ph}}{c^2} E_{ph}; V_{ph} = \frac{c}{n}; n > 1; V_{ph} = nc; n < 1$$

The left side of equation (65) is a strictly negative $A < 0$ for any differences in the initial and final velocity $\vec{V}_1 \neq \vec{V}_0$ of the electron and is zero $A = 0$ only if, after the emission of a photon by the medium, the final velocity of the electron exactly coincides with the initial velocity $\vec{V}_1 = \vec{V}_0$.

However, this is possible only for ideal conservative systems in which there is no energy dissipation at all. In the considered coordinate system, in the initial state, the electron has zero velocity, and in the final state, an electron having any velocity other than zero will have a higher energy in comparison with the energy of a stationary electron in the initial state.

Therefore, it follows from the negativity of the left side of equation (65) that for the generation of a positive-energy $\varepsilon_{ph} > 0$ VCR photon by the medium, the negativity condition of the right side of equation (65) must be fulfilled, which has the form of inequality $B < 0$.

Thus, a necessary and sufficient condition for the emission of a photon by a medium is the fulfillment of a strict inequality $B < 0$, which leads to the fulfillment of the condition for the implementation of the VCR (2), where it is the sign of strict inequality that must be considered. Due to this, condition (2) is in accordance with the experimental data of Cherenkov [31], shown in Fig.1, according to which the effect of the VCR is observed in the entire range of angles $0 \leq \theta < \theta_m$ from zero to the maximum angle of solution of the cone of the VCR, that is, at all angles.

In [11], an analogy was noted between the obtained condition for the implementation of the VCR (2) and the condition for the implementation of the anomalous Doppler effect, as well as the condition for dissipative instability generating the generation of waves with negative energy, as the Landau criteria (62) [93], [94]. In this case, unlike the anomalous Doppler effect, the emission of a microwave radiation by the medium when the left side (65) is negative is associated with an increase in the energy of the electron, and not with the transition of the emitting system with internal degrees of freedom to an excited state. On the other hand, if we consider the system as a whole as consisting of a moving medium and a stationary electron, then it is the electron that is assigned the role of an internal degree of freedom, the excitation of which is carried out precisely as a result of the emission of the VCR quantum by the medium.

## 6.3 The VCR in the interaction of a relativistic charged particle with a laser beam and with cosmic microwave background radiation

The possibility of implementing the VCR when a relativistic charged particle interacts with a laser beam [95] or with cosmic background background radiation [96] was previously considered on the basis of the well-known theory of the Frank-Tamm and Ginzburg IVF [38]. At the same time, it was shown that with the available energies of relativistic particles, there is no such possibility in both cases.

In both cases, the opposite conclusion was obtained within the framework of the new theory of microwave radiation, despite the fact that it is based on the same initial data from [95] and [96] on the level of refractive index fluctuations in a medium consisting of photons from a laser beam or cosmic background microwave radiation.

In particular, the following estimate of the refractive index of a medium of laser radiation photons was used, obtained in [95] for an optical laser:

$$n = 1 + \Delta n \qquad (66)$$

$$\Delta n = \frac{14\alpha E^2}{45\pi E_0^2} \approx 6.5 \times 10^{-11}, \qquad (67)$$

$$E_0 = \frac{m_e c^2}{e\lambda_c} = \frac{m_e^2 c^3}{e\hbar} \approx 1.3 \times 10^{18} \frac{V}{m};$$

The numerical estimate of the refractive index of the medium indicated in (65), (67) corresponds to the magnitude of the electric field strength $\frac{E}{E_0} = 3 \times 10^{-4}$ of the laser radiation, which is associated with the peak radiation power $S = 3 \times 10^{22} \frac{W}{sm^2}$ at the next photon energy limitation $\hbar\omega < 12 GeV$.

Let us consider the conditions of the VCR in the interaction of a relativistic charged particle with a medium, when, in the general case, the medium polarized by a relativistic particle is characterized by an extremely small difference from unity for the refractive index of the medium. In this case, conditions (58) or (2) of the VCR has the form:

$$V_0 > V_{th} = \frac{c}{n_*}; \gamma = \frac{1}{\sqrt{1 - \frac{V_0^2}{c^2}}} > \gamma_{th} = \frac{1}{\sqrt{1 - \frac{V_{th}^2}{c^2}}};$$

$$0 \leq \theta = \theta_{max} \frac{1}{2}\sqrt{1 - \frac{\gamma_{th}^2}{\gamma^2}} \leq \theta_{max} = \frac{1}{\gamma_{th}} \ll 1; \quad (68)$$

$$n = 1 + \Delta n; \Delta n \ll 1;$$

$$n_* = n + \sqrt{n^2 - 1} \approx 1 + \sqrt{2\Delta n} + \Delta n + O((\Delta n)^2)$$

According to the well-known condition for the implementation of the IVF, given in (1), the threshold value of the parameter $\gamma$ in the Tamm-Frank and Ginzburg theory, taking into account (66), (67), has the value:

$$\gamma_{th} \cong \frac{1}{\sqrt{2\Delta n}} \approx 8.8 \times 10^4; \Delta n = 6.5 \times 10^{-11} \quad (69)$$

The estimate given in (69) corresponds to the proton energy $E_p^{th} \approx 83 TeV$. Such energy is still unattainable even with the most modern LHC accelerator [13].

If the relativistic particle interacting with the laser beam is an electron, then the energy $E_e > 45 GeV$ of the electron is needed to realize the microwave in such an interaction, which is also unattainable under normal laboratory conditions.

On the other hand, from the condition for the emission of the VCR by the medium presented in (68), taking into account the same estimates (66), (67) to distinguish the refractive index from unity, the new VCR theory gives an estimate of the threshold value, which is almost 400 times less than in (69) and has the form [13]:

$$\gamma_{th} \cong \frac{1}{(8\Delta n)^{1/4}} \approx 2.09 \times 10^2; \Delta n = 6.5 \times 10^{-11} \quad (70)$$

According to (70), the proton energy $E_{th}^p = 209.4 GeV$ or the electron energy $E_e > 0.1047 GeV = 104.7 MeV$, which are already quite achievable in modern accelerators, is already sufficient for the implementation of the VCR.

Figure 3 shows a comparison of the estimates of the maximum angle of the VCR cone, which corresponds to estimates (69) and (70) for the threshold velocity of a charged particle interacting with a laser beam.

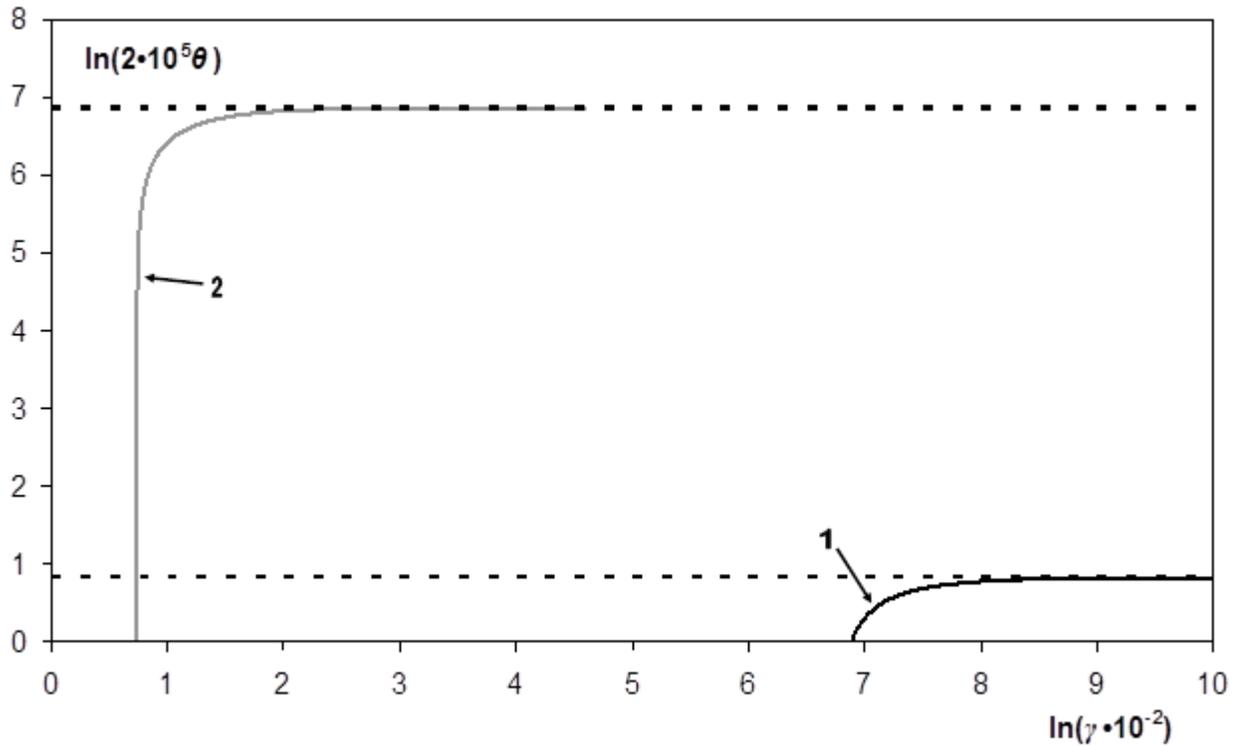

Fig.3 The angle of the VCR cone during the interaction of a charged particle with a laser beam $\theta = \theta_{max}\sqrt{1-(\frac{\gamma_{th}}{\gamma})^2}$; $\theta_{max} = \frac{1}{\gamma_{th}}$; graph of the dependence $\ln(2\times 10^5 \theta)$ on $\ln(\gamma \times 10^{-2})$

1. The quantum VCR theory [95], based on the theory of Minkowski (I.M. Dremin, 2002): $\theta_{max} = 1.14\times 10^{-5}$; $\gamma_{th} = 8.8\times 10^4$; $n = 1+\Delta n, \Delta n = 6.5\times 10^{-11}$

2. The quantum VCR theory [13], based on the theory of Abraham: $\theta_{max} = 4.78\times 10^{-3}$; $\gamma_{th} = 2.09\times 10^2$; $n = 1+\Delta n, \Delta n = 6.5\times 10^{-11}$

In [13], a similar consideration was carried out in connection with the problem of the interaction of fast charged particles of cosmic rays with cosmic background microwave radiation (CMB), investigated in [96].

At the same time, according to [96], based on the estimate (69) following from the VCR theory of the Tamm-Frank and Ginzburg [22], [23], [38], in the modern era, at the observed energy density of the background microwave radiation (when $n-1 \approx 10^{-42}$) it is not possible to implement the VCR as a result of the indicated interaction of cosmic rays with CMB at a threshold value $\gamma_{th} \cong 10^{21}$.

As shown in [13] from the VCR condition (70) given by the new quantum VCR theory [11]-[13], even with such extremely small deviations of the refractive index of the medium from unity, it is possible to implement VCR in the modern era at a threshold value $\gamma_{th} \cong 10^{11}$.

At the same time, the difference in estimates given in the new and generally accepted VCR theories for the threshold energy of the VCR realization in the CMB environment is no longer two orders of magnitude, as in Fig.2, but ten orders of magnitude different from each other [13].

Thus, a significant difference has been obtained in the quantitative estimates of the VCR theories based on the theory of Minkowski and the theory of Abraham. It is related to the consideration of the microscopic mechanism of the emission of a photon by the medium, due to which, instead of the refractive index $n$ appearing in the well-known condition for the implementation of the VCR (1), the new theory of the microwave and condition (2) uses the value $n_*$.

Therefore, it is precisely with a small difference in the refractive index $n$ from unity that the value $\Delta n = n-1$ in (1) and (69) can differ by many orders of magnitude from the value $n_* - 1 \approx \sqrt{2\Delta n} \gg \Delta n$ included in condition (2) and (68) in the limit $\Delta n \ll 1$.

## 6.4 The application of the new VCR theory for particle identification and in accelerator protection systems

The condition for the implementation of the VCR effect in the form (2), obtained in the new quantum VCR theory based on Abraham's theory, as shown in the previous sub-chapter, can lead to estimates of the threshold for the implementation of VCR (70), which differs by many orders of magnitude from the estimate (69) obtained from condition (1) of the well-known VCR theory of Tamm-Frank and Ginzburg.

Such a significant difference is due to the consideration in (2) and (70) of the microscopic mechanism of the emission of the VCR by the medium, which is not taken into account in the known condition (1), when considering precisely the case of a medium with a close to unity refractive index.

Therefore, it is of interest to consider the difference in the application of the conditions of the RFI in the form of (1) and (2) in connection with the analysis of data obtained using Cherenkov detectors for the identification of high-energy relativistic particles, which usually use a medium with such a close to unity refractive index.

Indeed, it is a medium with a refractive index close to unity that is used in these detectors in order to increase its sensitivity when identifying relativistic particles with high energy. [17], [18], [97]-[99].

In particular, to identify particles with energies exceeding several GeV, materials with a refractive index close to unity are required when large gas chambers are used. In new particle detection and identification systems, the refractive index can also be adjusted to various values close to unity by the angle of the cone [97]-[99].

Thus, although higher refractive indices lead to lower VCR thresholds and higher photon output, they are not always desirable for Cherenkov detectors. The reason is that a high refractive index reduces the sensitivity of the VCR cone angle to small changes in the velocity of the detected particle.

Ring imaging of RICH-type is known, where Cherenkov radiators are used to identify unknown elementary particles by determining their mass from the angle of the VCR cone that is emitted when particles pass through a detector device. However, as the momentum of the particles increases, the angle of inclination of the cone increases only to a value determined only by the refractive index and then does not actually change with increasing momentum of the particles, which makes it difficult to effectively distinguish between different particles.

For example, quartz has a refractive index of about 1.4, and its corresponding pulse coverage for the identification of pions and kaons is currently limited to pulses below 6 GeV/s.

Increasingly high-energy particles with pulses exceeding this value emit light in a small range of angles of the solution of the cone [44.30°, 44.42°], regardless of the velocity and type of particles. In this scenario, measuring the angle of the solution of the cone of the VCR cannot lead to the identification of the corresponding type of particles [97]- [99].

In this regard, for example, metal-based anisotropic meta-materials with one component of the effective refractive index close to unity [99]-[102] have been proposed.

The sensitivity value is usually calculated for Cherenkov detectors based on condition (1) of the well-known Tamm-Frank and Ginzburg VCR theory [38] and has the form [2] (see (7.15) in the book [2]):

$$\frac{d\theta_C}{dV} = \frac{n}{c}\frac{\cos^2\theta_C}{\sin\theta_C};$$
$$\cos\theta_C = \frac{1}{\beta n}; n > 1 \qquad (71)$$

Similarly, from condition (2) of the new VCR theory [11]-[13], instead of (71), we obtain the following representation for the sensitivity value:

$$\frac{d\theta_m}{dV} = \frac{n_*}{c}\frac{\cos^2\theta_m}{\sin\theta_m};$$
$$\cos\theta_m = \frac{1}{\beta n_*};$$
$$n_* = n + \sqrt{n^2 - 1}, n > 1; n_* = \frac{n}{1 - \sqrt{1-n^2}}, n < 1 \qquad (72)$$

The difference between (72) and (71) is that the magnitude of the angle $\theta = \theta_C = \arccos(1/\beta n)$ in (1) and (71) corresponds to the angular position of the maximum in the angular distribution of the intensity of the VCR inside the observed (see Fig.1) VCR cone $\theta_C < \theta_m$, where $\theta_m = \arccos(1/\beta n_*)$ according to the threshold condition (2).

The value $n_*$ included in (72) also always exceeds the value of the refractive index $n$.

As noted above, due to the problem of the accuracy of particle identification when using Cherenkov detectors, the case of refractive indices close to unity is of particular importance. In this regard, we compare estimates (71) and (72) in the limit $n = 1 + \Delta n; \Delta n \equiv \delta \ll 1$, also considered in (68). With this limit from (71) for ultra-relativistic particles $\beta \to 1$, we obtain an estimate:

$$S_C = \frac{d\theta_C}{d\beta} = \frac{1}{\beta\sqrt{n^2\beta^2 - 1}} \cong \frac{1 + O\left(\frac{1-\beta}{2\delta\beta^2}\right)}{\beta^2\sqrt{2\delta}} \gg 1; \qquad (73)$$
$$\delta = n - 1 \ll 1; \frac{1-\beta}{2\delta\beta^2} \ll 1$$

In the same limit, from (72) we obtain an estimate in the form:

$$S_m = \frac{d\theta_m}{d\beta} = \frac{1}{\beta\sqrt{n_*^2\beta^2 - 1}} \cong \frac{1 + O\left(\frac{1-\beta}{2\beta^2\sqrt{2\delta}}\right)}{\beta^2(8\delta)^{1/4}} \gg 1; \quad (74)$$

$$\delta = n - 1 \ll 1; \frac{1-\beta}{2\beta^2\sqrt{2\delta}} \ll 1$$

From the comparison of (74) and (73) it follows that in the considered limit there is a relation in the form:

$$\frac{S_C}{S_m} \cong \left(\frac{2}{n-1}\right)^{1/4} \quad (75)$$

Thus, from the above estimates (71)-(75), it follows that it is important to take into account the difference between the well-known theory of the Tamm-Frank and Ginzburg theory [22],[23], [38] and a new microscopic theory of the VCR emission by medium [11]-[13] to increase the efficiency of particle identification using Cherenkov detectors.

Another example of a problem in which it is also important to take into account the differences in the VCR conditions (1) and (2) is related to the modern development of protection systems for powerful accelerators using the VCR effect for damage detection.

In particular, [20] proposed a new type of equipment protection system based on optical fibers, which will complement existing systems that improve existing characteristics. These fibers are laid along the accelerator beam line with a length of ~100 m, providing continuous coverage over this distance. When relativistic particles pass through these fibers, they generate Cherenkov radiation in the optical spectrum. This radiation propagates in both directions along the fiber and can be detected at both ends [20]. The calibration-based method makes it possible to determine the location of the Cherenkov radiation source with an accuracy of 0.5 m with a resolution of 1 m. [20]

In most of the applications under consideration, the VCR sensor will detect secondary particles rather than primary ones, that is, a stream of charged particles that occurs when an energetic particle collides with a surface. This release of charged particles, usually electrons, occurs as a result of some external event, it can be a loss of a beam or a radio frequency breakdown. The flow of charges propagates in free space to the optical fiber, where the fiber intersects at a small solid angle. This solid angle depends on the diameter of the fiber core. Wider fibers provide a more powerful signal because more particles in the stream interact with the fiber.

However, this larger solid angle also increases the duration of the VCR pulse and limits the resolution of the device [103], [104].

For example, a beam with lower energy will generate fewer charged particles in case of signal loss than a beam with higher energy, which means that a larger diameter fiber will be required to collect enough signal for analysis. As soon as the particles enter the fiber, the detection process begins. Charged particles pass through the fiber relatively without being absorbed and can generate Cherenkov radiation.

To date, the well-known VCR condition in the form (1) has been used to evaluate the possibility of such generation and the angle of the cone of the VCR.

For example, for silica optical fibers (n = 1.46), the threshold electron velocity $\beta_{th} = V/c \approx 0.69$ is obtained from condition (1). For all electrons with a velocity $\beta \geq 0.69$ in the signal optical fiber, the VCR will be generated. The corresponding threshold energy of the electron will be 186 keV. The subsequent waveguide propagation of the VCR will be determined by the angle $\theta_C = \arccos(1/n\beta_{th}) \approx 6.95^0$ of the VCR cone, determined from (1). Based on the new VCR theory and the conditions for the implementation of VCR (2), the following estimates can be obtained for comparison. With the specified value of the refractive index $n = 1.46$ the value of the effective refractive index $n_* = n + \sqrt{n^2 - 1} \approx 2.52$ is obtained, taking into account the well-known [1]-[4] microscopic mechanism of photon emission by a medium, that is, an optical fiber polarized by a fast electron. At the same time, according to (2), the threshold value of the electron velocity turns out to be equal $\beta = \beta_{th}^A \approx 0.4$, which is noticeably lower than the threshold of the velocity obtained above based on condition (1) of the well-known theory of the VCR. The corresponding value of the threshold electron energy will be approximately 46 keV, which is almost four times lower than the threshold electron energy given above and obtained on the basis of the VCR condition in the form of (1). An even more significant difference in the estimate of the angle in the solution of the VCR cone is obtained by comparing the above value of about 7 degrees and the value $\theta_m = \arccos(1/n_*\beta) \approx 55^0$ obtained using such the same speed values $\beta = 0.69$ as in the above estimate $\theta_c$. However, when using the new threshold value $\beta = \beta_{th}^A \approx 0.4$, a close estimate $\theta_m = \arccos(1/n_*\beta_{th}^A) \approx 7.22^0$ is obtained.

Thus, examples of the application of the new VCR theory [11]-[13], based specifically on the theory of Abraham, are given, which allows us to describe the well-known [1]-[4] microscopic mechanism of emission of the VCR by a shock-polarized medium, which was not previously taken into account in the well-known VCR theory of Tamm-Frank and Ginzburg [22], [23], [38].

## 7. Appendix A: The Abraham theory and the Abraham force

Maxwell's equations in the medium have the form [38], [41]:

$$\frac{1}{c}\frac{\partial \vec{D}}{\partial t} + \frac{4\pi}{c}\vec{j}_{ext} = rot\vec{H} \qquad (A1)$$

$$-\frac{1}{c}\frac{\partial \vec{B}}{\partial t} = rot\vec{E} \qquad (A2)$$

$$div\vec{D} = 4\pi\rho_{ext} \qquad (A3)$$

$$div\vec{B} = 0 \qquad (A4)$$

$$\vec{D} = \vec{E} + 4\pi\vec{P} \qquad (A5)$$

$$\vec{B} = \vec{H} + 4\pi\vec{M} \qquad (A6)$$

The law of conservation of electromagnetic field energy follows from these equations:

$$\frac{1}{8\pi}\frac{\partial}{\partial t}\left((\vec{D}\vec{E}) + (\vec{B}\vec{H})\right) = -(\vec{j}_{ext}\vec{E}) - div\vec{S}; \qquad (A7)$$

$$\vec{S} = \frac{c}{4\pi}\left[\vec{E}\times\vec{H}\right] \tag{A8}$$

The momentum density of the field in the form of Abraham and in the form of Minkowski:

$$\vec{g}_A = \vec{S}/c^2 = \frac{1}{4\pi c}\left[\vec{E}\times\vec{H}\right];$$
$$\vec{g}_M = \frac{1}{4\pi c}\left[\vec{D}\times\vec{B}\right] \tag{A9}$$

The relation follows directly from Maxwell's equations (A1)-(A6):

$$-\frac{\partial \vec{g}^A}{\partial t} + \vec{Q} = \vec{F}^L + \vec{F}_D^A + \vec{F}_B^A;$$

$$\vec{F}^L = \rho_{ext}\vec{E} + \frac{1}{c}\left[\vec{j}_{ext}\times\vec{B}\right]; \vec{F}_D^A = \frac{1}{c}\frac{\partial}{\partial t}\left[\vec{P}\times\vec{H}\right]; \vec{F}_B^A = \frac{1}{c}\frac{\partial}{\partial t}\left(\left[\vec{E}\times\vec{M}\right] + 4\pi\left[\vec{P}\times\vec{M}\right]\right); \tag{A10}$$

$$Q_i = \frac{1}{4\pi}\left[E_i div\vec{D} + D_j\left(\frac{\partial E_i}{\partial x_j} - \frac{\partial E_j}{\partial x_i}\right) + B_j\left(\frac{\partial H_i}{\partial x_j} - \frac{\partial H_j}{\partial x_i}\right)\right]$$

Equation (A10) is obtained by vector multiplication from the left of equation (A1) by a vector $\vec{B} = \vec{H} + 4\pi\vec{M}$ and its subsequent addition with equation (A2), after vector multiplication of Eq. (A2) from the left by a vector $\vec{D} = \vec{E} + 4\pi\vec{P}$. Then add a term $\rho_{exp}\vec{E}$ to the left and right sides of the resulting ratio, which is transformed in the left part of this ratio taking into account equation (A3).

In equation (A10), the vector $\vec{Q}$ cannot generally be represented as the divergence of a second-rank tensor. Therefore, we transform equation (A10) to a form satisfying the integral law of conservation of momentum of the electromagnetic field. At the same time, we take into account that there is an identical representation:

$$Q_i + (F_D^{AM})_i + (F_B^{AM})_i = \frac{\partial \sigma_{ij}}{\partial x_j};$$

$$\sigma_{ij}^A = \frac{1}{8\pi}\left[E_i D_j + E_{j+H} D_i + H_i B_j + H_j B_i - \delta_{ij}\left((\vec{E}\vec{D}) + (\vec{H}\vec{B})\right)\right];$$

$$\vec{F}_D^{AM} = \frac{1}{2}rot\left[\vec{P}\times\vec{E}\right] + \vec{F}_D; (F_D)_i = \frac{1}{2}\left(P_j\frac{\partial E_j}{\partial x_i} - E_j\frac{\partial P_j}{\partial x_i}\right); \tag{A11}$$

$$\vec{F}_B^{AM} = \frac{1}{2}rot\left[\vec{M}\times\vec{H}\right] + \vec{F}_B; (F_B)_i = \frac{1}{2}\left(M_j\frac{\partial H_j}{\partial x_i} - H_j\frac{\partial M_j}{\partial x_i}\right)$$

As a result, when the sum of terms $\vec{F}_D^{AM} + \vec{F}_B^{AM}$ is added to the left and right sides of equation (A10), the law of conservation of momentum of the electromagnetic field is obtained in the form proposed by Abraham:

$$-\frac{\partial g_i^A}{\partial t} + \frac{\partial \sigma_{ij}}{\partial x_j} = F_i^L + F_i^A; i = 1,2,3;$$ (A12)

$$\vec{F}_{Abr} = \vec{F}_D^A + \vec{F}_D^{AM} + \vec{F}_B^A + \vec{F}_B^{AM}$$

For simplicity, we will consider the case of a non-magnetic medium $\vec{B} = \vec{H}$. At the same time, in (A12), for the Abraham force, we have the following representations:

$$\vec{F}_{Abr} = \vec{F}_D^A + \vec{F}_D^{AM};$$

$$\vec{F}_D^A = \frac{1}{c}\frac{\partial}{\partial t}\left[\vec{P} \times \vec{H}\right];$$ (A13)

$$\vec{F}_D^{AM} = \frac{1}{2} rot\left[\vec{P} \times \vec{E}\right] + \frac{1}{2}\left(P_j \frac{\partial E_j}{\partial x_i} - E_j \frac{\partial P_j}{\partial x_i}\right)$$

In particular, for an isotropic medium characterized by permittivity $\varepsilon = \varepsilon(\vec{x};t)$, under the condition of $\vec{D} = \varepsilon(x,t)\vec{E}$ from the Eq. (A13) we obtain the most common representation for the sum of the densities of the Abraham force (the first term in (A14)) and the Abraham-Minkowski force (the second term in (A.14), which is also present in the Minkowski theory [5], [38]):

$$\vec{F}_{Abr} = \frac{(\varepsilon - 1)}{4\pi c}\frac{\partial}{\partial t}\left[\vec{E} \times \vec{H}\right] - \frac{\vec{E}^2}{8\pi}\vec{\nabla}\varepsilon;$$ (A14)

$$\varepsilon = n^2$$

B (A14) is the refractive index of the medium. For a homogeneous medium, the second term in (A14), which is also present in Minkowski theory [5], [38], vanishes when we obtain the representation used in the main text for the Abraham force density $\vec{f}^A = \vec{F}_{Abr}(\varepsilon = const)$ in (27).

## 8. Appendix B. Abraham's force impulse

The derivation of relations (14) and (16), following the work of D. V. Skobeltsyn (1977) [26] is represented.

In [26], the representation for the Abraham force momentum (14) and the following relation (15) are derived only for the case $n > 1$, which is used in [38], [42], [43] to substantiate the idea of the "usefulness" of applying Minkowski theory in the quantum theory of the Ginzburg quantum theory [23]-[25].

As in [26], we take the direction of the VCR beam as the direction of the axis $x$ and use the following representation to describe the change in time and space for the corresponding EMF energy density :

$$w = u\left(v\left(t - \frac{x}{c_*}\right)\right); \tag{B.1}$$

$$c_* = \frac{c}{n}, n > 1; c_* = cn, n < 1$$

B (B.1) $t$ - is the time elapsed after the beginning of the emission of a plume of the VCR waves emitted by an VCR source that is stationary relative to an equally stationary medium [26].

Such an examination in [26] actually corresponds to the microscopic mechanism of the emission of the VCR by a polarized medium, rather than directly by a rapidly moving charged particle [1]-[4]. However, in [26] did not pay attention to this circumstance, and the criticism of Minkowski theory [5] and the quantum VCR theory of the Ginzburg based on it [23], [42], contained in [26], on the contrary, proceeded from the idea that the emitter of the VCR is a directly charged particle, as it is and it is considered in [23].

For the Abraham force density modulus in [26] in the case of a plane wave field, the representation is used (see (1.8) in [26] for the value $n > 1$ and the footnote on page 321 in [38] for the value $n < 1$):

$$f^A = (n^2 - 1)\frac{\partial}{\partial t}g^A; g^A = \frac{\left[\vec{E} \times \vec{H}\right]}{4\pi c} \tag{B.2}$$

$$g^A = \frac{w}{c^2}V_g; V_g = c/n, n > 1; V_g = cn, n < 1$$

From (B.2), taking into account the relation $\partial w/\partial t = -c_*\partial w/\partial x$, following from (B.1), we obtain the representation:

$$f^A = -(n^2 - 1)\frac{c_*^2}{c^2}\frac{\partial w}{\partial x} \tag{B.3}$$

At the leading edge of the VCR wave, in the transition zone, the magnitude of the Abragm force density, averaged over a long period of time compared to the oscillation period of the wave, is nonzero and gives the resulting force acting on the medium (per unit cross-sectional area of the wave train), which is equal to:

$$P = \int_x^\infty dx f^A = (n^2 - 1)\frac{c_*^2}{c^2}w \tag{B.4}$$

The energy of the wave-packet corresponding to the photon energy $E_{ph} = wL$ for the wave-packet with a unit cross-sectional area is determined by the length $L = c_*T$ of the wave-packet, where $T$ - is the time after which the wave-packet detaches from the VCR source. Therefore, from (B.4) we obtain the following representation for the momentum of the Abraham force $F^A = PT$, expressed in terms of the photon energy and having the form:

$$F^A = (n^2 - 1)\frac{c_*}{c^2}E_{ph} \tag{B.5}$$

The representation (B.5) coincides with (14) for the refractive index $n>1$, but (B.5) also follows to the representation (16) for the Abraham force pulse given in the main text when $n<1$.

## 9. Conclusions

Thus, in this paper it is shown that the VCR theory, based on the theory of Abraham, more accurately describes the observed threshold of the VCR effect than the theory of VCR, based on the theory of Minkowski and therefore does not take into account the known microscopic mechanism of emission of the VCR by a polarized medium. This, in turn, allows us to conclude that the still popular idea of the usefulness of using Minkowski theory in the theory of electromagnetic radiation, which continues to be widely used when considering the effects of electromagnetic radiation in the environment, despite its refutation by experiments on direct measurement of the Abraham force, is unfounded. [6]-[8], [45], [46]. This well-known view is refuted by the example obtained in [11]-[13] of the use of Abraham's theory in the creation of the theory of the VCR, consistent with the observational data.

This removes an important basis for the ongoing discussion of the dilemma of Minkowski and Abraham's theories for more than a century, and finally makes it possible to use Abraham's theory in describing EMF in the environment and the phenomena of interaction of EMF and matter.

In addition, this paper considers possible areas of application of the new quantum VCR theory [11]-[13], the conclusions of which differ significantly from the conclusions of the quantum VCR theory of Ginzburg [4], [23], [38], especially in the case of media with a refractive index close to unity. This difference is of practical importance in the analysis of data from Cherenkov detectors and in the design of protection systems for modern accelerators. Of fundamental interest is also the use of a new VCR theory based on the theory of Abraham and in connection with the problem of observing VCR in the interaction of relativistic charged particles with photonic gas or a laser beam.

The application of the concept of the finite real mass of a photon in a medium, which follows from Abraham's theory, provides new insights into a number of problems of nonlinear optics. It is shown that the violation of spherical symmetry in the propagation of EMF waves leads to the appearance of a photon mass (43) in the waveguide comparable to the photon mass (6) in Abraham's theory.

In particular, this applies to the problems of energetically efficient propagation of electromagnetic waves in a medium in the form of practically diffraction-free Bessel rays and associated Cherenkov-type radiation [74-79]. Such terahertz radiation, which has numerous applications, can already be adequately described within the framework of the new VCR theory based on Abraham's theory, since it has so far been inaccessible to understanding within the framework of the generally accepted VCR theory based on Minkowski theory. [77]-[79], [83]-[88].

6243. Brevik I, Experiments in phenomenological electrodynamics and the electromagnetic energy-momentum tensor, Phys. Rep., 52, 133-201 (1979); https://doi.org/10.1016/0370-1573(79)90074-7
44. Toptygin I. N., Quantum field description in macroscopic electrodynamics and properties of photons in transparent media, UFN, 187, 1007 (2017)
45. Rikken G L J A, and van Tiggelen B A, Measurement of the Abraham force and its predicted QED corrections in crossed electric and magnetic fields, Phys. Rev. Lett., 107, 170401 (2011); https://doi.org/10.1103/PhysRevLett.107.170401
46. Rikken D L J A, and van Tiggelen B A, Observation of the intrinsic Abraham force in time-varying magnetic and electric fields, Phys. Rev. Lett., 108, 230402 (2012); https://doi.org/10.1103/PhysRevLett.108.230402
47. Astrath N G C, Malcarne L C, Baesso M L, Lukasievicz G V B, and Bialkowski S E, Unravelling the effects of radiation forces in water, Nature Comm., 5, 4363 (2014); https://doi.org/10.1038/ncomms5363
48. Zhang L, She W, Peng N, and Leonhardt U, Experimental evidence for Abraham pressure of light, New J. Phys., 17, 053035 (2015); https://doi.org/10.1088/1367-2630/17/5/053035
49. Kundu A, Rani R, and Hazra K S, Graphen oxide demonstrates experimental confirmation of Abraham pressure on solid surface, Sci. Rep., 7, 52538 (2017); https://doi.org/10.1038/srep42538
50. Choi H, Park M, Elliot D S, and Oh K, Optomechanical measurement of the Abraham force in an adiabatic liquid-core optical-fiber waveguide, Phys. Rev. A, 95, 053817 (2017); https://doi.org/10.1103/PhysRevA.95.053817
51. She W, Yu J, and Feng R, Observation of a push force on the end face of a nanometer silica filament by outgoing light, Phys. Rev. Lett., 101, 243601 (2008); https://doi.org/10.1103/PhysRevLett.101.243601
52. She W, Yu J, and Feng R, Reply to the Comment by Iver Brevik, Phys. Rev. Lett., 103, 219302 (2009); https://doi.org/10.1103/PhysRevLett.103.219302
53. Conde J G L, and Saldanha P L, Recoil momentum of an atom absorbing light in a gaseous medium and Abraham-Minkowski debate, Phys. Rev. A, 108, 013511 (2023); https://doi.org/10.1103/PhysRevA.108.013511 ; arXiv: 2307.06938v1 [physics. optics] 15 Jul 2023
54. Burt M G, Peierls R E, The momentum of a light wave in a refracting medium, Proc. Roy. Soc. A 333, 149 (1973); https://doi.org/10.1098/rspa.1973.0053
55. Jones R V, Radiation pressure in a refractive medium, Nature, 167, 439 (1951); https://doi.org/10.1038/167439a0